\documentclass[%
 reprint,
superscriptaddress,
 amsmath,amssymb,
pra,
floatfix,
]{revtex4-1}

\usepackage{graphicx}
\usepackage{dcolumn}
\usepackage{bm}
\usepackage{braket}
\usepackage{float}
\usepackage{expl3}
\usepackage{color}
\usepackage{subfigure}
\usepackage[T1]{fontenc}

\begin{document}

\preprint{APS/123-QED}

\title{Near-term performance of quantum repeaters with imperfect ensemble-based quantum memories}

\author{Yufeng Wu}
\affiliation{Institute for Quantum Science and Technology, and Department of Physics and Astronomy, University of Calgary, Calgary T2N 1N4, Alberta, Canada}

\author{Jianlong Liu}
\affiliation{Hefei National Laboratory for Physical Sciences at Microscale and Department of Modern Physics, University of Science and Technology of China, Hefei, Anhui 230026, China}
\affiliation{CAS Center for Excellence in Quantum Information and Quantum Physics, University of Science and Technology of China, Hefei, Anhui 230026, China}
\author{Christoph Simon}%
\affiliation{Institute for Quantum Science and Technology, and Department of Physics and Astronomy, University of Calgary, Calgary T2N 1N4, Alberta, Canada}




\date{\today}

\begin{abstract}
  We study the feasibility of meaningful proof-of-principle demonstrations of several quantum repeater protocols with photon (single-photon and photon-pair) sources and atomic-ensemble based quantum memories. We take into account non-unit memory efficiencies that decay exponentially with time, which complicates the calculation of repeater rates. We discuss implementations based on quantum dots, parametric down-conversion, rare-earth-ion doped crystals, and Rydberg atoms. Our results provide guidance for the near-term implementation of long-distance quantum repeater demonstrations, suggesting that such demonstrations are within reach of current technology. 
  \end{abstract}

\maketitle


\section{\label{sec:1} Introduction}

The future quantum internet \cite{kimble2008quantum, simon2017towards, wehner2018quantum} is expected to enable many applications, including secure communication \cite{bennett2014quantum, ekert1991quantum, gisin2002quantum}, quantum-enhanced distributed sensing \cite{gottesman2012longer, komar2014quantum}, and distributed quantum computing \cite{beals2013efficient}. 
Despite the recent progress on satellite quantum communication \cite{liao2017satellite, ren2017ground}, quantum repeaters \cite{briegel1998quantum}, where entanglement is first generated and stored in a number of elementary links, followed by entanglement swapping steps to extend it to the total distance, are expected to be essential for the quantum internet to become a reality \cite{simon2017towards}. Our focus here is on so-called first-generation quantum repeaters \cite{muralidharan2016optimal} without quantum error correction, and in particular on quantum repeater protocols with atomic-ensemble based quantum memories \cite{Sangouard}, but we note that there is also a lot of recent work on near-term quantum repeaters with single quantum systems \cite{rozpkedek2019near,santra2019quantum,krutyanskiy2019light}.

There has recently been significant experimental progress in the entanglement of two remote quantum memories (corresponding to one repeater link) \cite{bao2012quantum, pfaff2014unconditional, vittorini2014entanglement, humphreys2018deterministic, yu2019entanglement}. For example, Ref. \cite{yu2019entanglement} demonstrated entanglement between two $^{87}\mathrm{Rb}$ atomic ensembles separated by 22 km of coiled fiber. Simple repeater demonstrations with two links are now being envisioned by various experimental groups. It is therefore important to make realistic theoretical predictions for such demonstrations.

Previous papers have studied the performance of repeaters with imperfect ensemble-based memories \cite{brask2008memory, Sangouard, PhysRevLett.113.053603, Krovi2016, collins2007multiplexed, shchukin2019waiting}, but they have typically either focused on more long-term scenarios and made assumptions that are not quite realistic yet, such as a high degree of multiplexing, or have made idealizations that may affect quantitative rate predictions, such as a simple cut-off for the storage time, rather than an exponential decay. Decoherence in ensemble-based memories results in a reduction of efficiency rather than fidelity \cite{staudt2007interference}. While this is positive from the point of achieving high final-state fidelity in quantum repeater protocols, it complicates the derivation of accurate repeater rates because it makes the swapping probabilities time-dependent. 

Here we focus on near-term repeaters with a small number of links, and where each node contains only the minimum necessary number of memories (one or two). We treat the effects of non-unit memory efficiency that decays exponentially in time, including the resulting time-dependence of the swapping probabilities. We evaluate the performance of four repeater schemes, namely that of Ref. 
\cite{PhysRevA.76.050301}, which uses single-photon sources and single-photon Bell measurements, that of Ref. \cite{PhysRevLett.113.053603}, which combines deterministic photon-pair sources with two-photon Bell-state measurements, as well as two schemes that combine non-deterministic photon-pair sources with single-photon \cite{PhysRevLett.98.190503} or two-photon \cite{Krovi2016} Bell-state measurements. 
We consider some promising implementations for each repeater scheme, such as quantum memory based on rare-earth-ion doped crystals \cite{siyushev2014coherent, xia2015all, fraval2005dynamic, longdell2005stopped, Zhong2015, Rancic2017, sabooni2010storage,usmani2010mapping,AMARI20101579,Bonarota_2011, timoney2012atomic, Jobez_2014,sabooni2013efficient} and Rydberg atoms \cite{PhysRevLett.87.037901, saffman2002creating, PhysRevA.97.053803, PhysRevLett.123.140504, yang2016efficient}, as well as photon sources based on SPDC \cite{Couteau_2018}, quantum dots \cite{senellart2017high, ding2016demand, wang2019demand}, and Rydberg atoms \cite{dudin887, dudin2012observation, li2019semi}, and we base our performance estimates for the different repeater schemes on the experimental status quo. 

The plan of the paper is as follows. In sec. \ref{sec:reps}, we introduce the four quantum repeater schemes that we focus on in this paper. In sec. \ref{sec:math}, we analyze the effect of imperfect quantum memories on repeater performance. In sec. \ref{sec:imp} we outline the most promising implementations. In sec. \ref{sec:numer}, we give numerical results for repeater rates under realistic conditions.

\section{Repeater protocols}
\label{sec:reps}

The basic principle of the quantum repeaters is to reduce the transmission loss by dividing the distance into small segments(link), where the entanglement is generated via Bell-state-measurement(BSM) in each segment and then extended via entanglement swapping \cite{Sangouard}.   In this paper, we focus on four repeater schemes, 
denoted using the form of ``a + b'', where ``a'' represents the type of photon sources and ``b'' represents the type of BSM. We list these repeater schemes as follows:

\begin{figure}[ht]
    \centering
    \includegraphics[width=\linewidth]{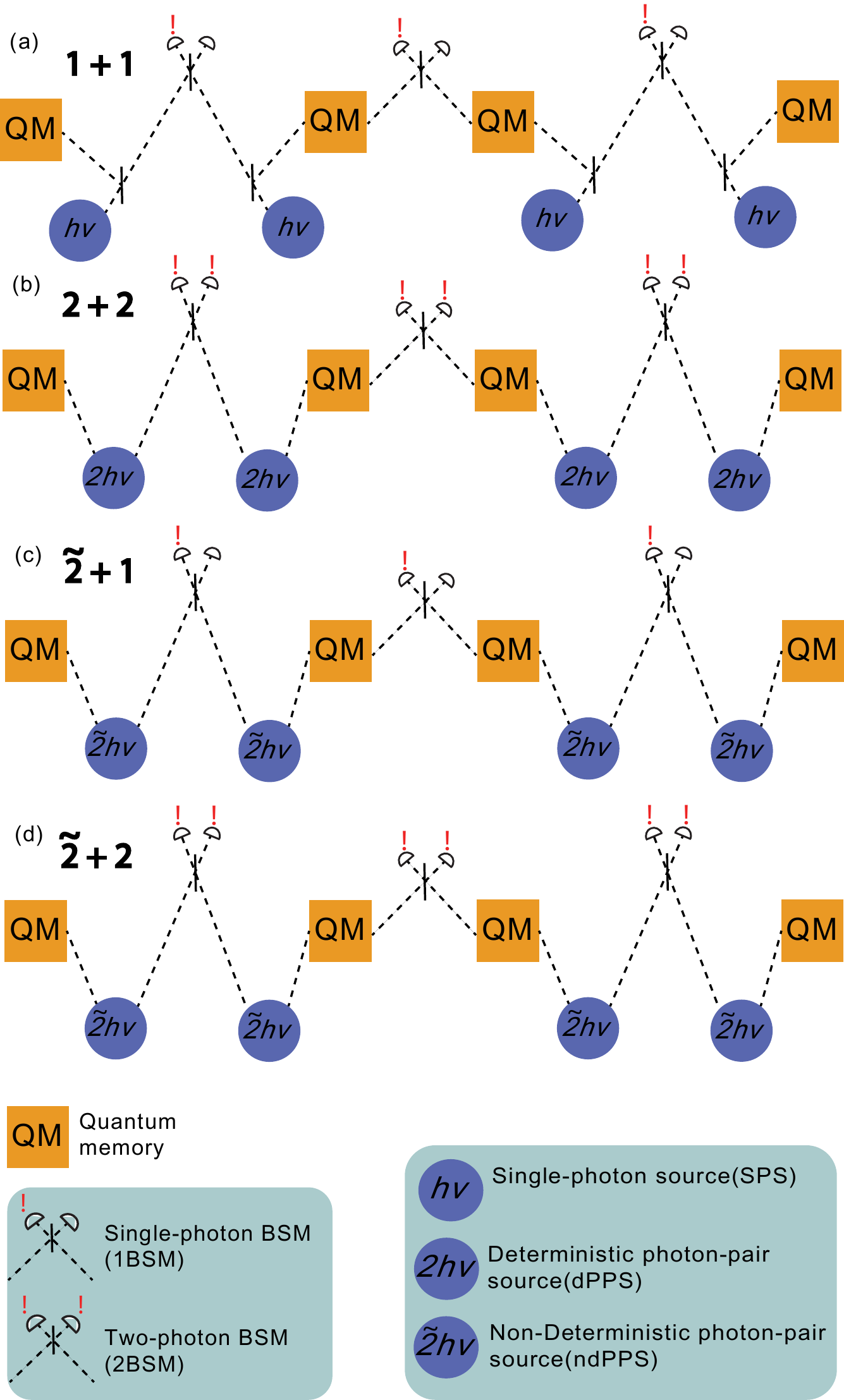}
    \caption{Sketches of quantum repeater protocols we mainly discuss in this paper. Here we show the two-link version of these repeater protocols. (a)Single-photon source(SPS) with single-photon BSM(``1 + 1''). (b)Deterministic photon-pair source(dPPS) with two-photon BSM(``2 + 2''). (c)Non-deterministic photon-pair source(ndPPS) with single-photon BSM(``$\tilde{2}$ + 1''). (d) Non-deterministic photon-pair source(ndPPS) with two-photon BSM(``$\tilde{2}$ + 2'').}
    \label{fig:rep_fig}
\end{figure}

\begin{itemize}
    \item 1 + 1: single-photon source(SPS) with single-photon BSM scheme \cite{PhysRevA.76.050301} is shown in Fig. \ref{fig:rep_fig}(a). The ``1 + 1'' protocol uses deterministic single-photon sources at each node. A single photon is generated in each node and sent through a local beam-splitter with probability $\gamma$ stored in a quantum memory and $1 - \gamma$ injected to the central interferometer. As the central beam splitter erases the which-path information, the successful detection of a single-photon in the central station will create an entangled state of two memories in the Fock space. 
    
    \item 2 + 2: deterministic photon-pair source(PPS) with two-photon BSM scheme \cite{PhysRevLett.113.053603} is shown in Fig. \ref{fig:rep_fig}(b). The protocol uses deterministic entangled photon-pair sources that emit a photon-pair sending to quantum memory and central station, respectively. Depending on the qubit encoding(time-bin, frequency, polarization, etc), the two-photon BSM will project the two quantum memories into the entangled state in the corresponding space. 
    
    \item $\tilde{2}$ + 1: non-deterministic photon-pair source(ndPPS) with single-photon BSM scheme \cite{PhysRevLett.98.190503} is shown in Fig. \ref{fig:rep_fig}(c). The ndPPS emits photon-pair probabilistically, with one photon stored in the quantum memory and another sent to the central station. As for the ``1 + 1'' scheme, a successful single-photon BSM will project the two memories into the entangled state in the Fock space. Similar to the DLCZ scheme \cite{duan2001long}, the emission probability should be small to suppress multi-pair emission.
    
    \item $\tilde{2}$ + 2: non-deterministic photon-pair source(ndPPS) with two-photon BSM scheme \cite{zhao2007robust,Krovi2016} is shown in Fig. \ref{fig:rep_fig}(d). This scheme is similar to the ``2 + 2'' scheme while using non-deterministic entangled photon-pair sources. It is possible, albeit with small probability, to get coincident photon-pair emission from two ndPPS. The successful two-photon BSM will project the two memories into the two-photon entangled state. Again the multi-pair emission probability has to be kept small; the associated errors can be mitigated by appropriately designed entanglement swapping \cite{zhao2007robust} and photon-number resolving detection \cite{Krovi2016}.
\end{itemize}

It is important to notice the different requirements for different repeater schemes. 
To begin with, single-photon BSM requires encoding qubits in the Fock space, which cannot be used directly for quantum communication tasks, and thus postselection is necessary to convert the qubits into useful two-photon entanglement state analogous to qubit states in two-photon BSM \cite{Sangouard}. Besides, single-photon BSM requires phase stability, while the phase is automatically stabled in two-photon BSM. On the other hand, for the scheme with SPS, the frequency should match both memory and telecom-band, in comparison to schemes with dPPS and ndPPS, where the two photons from a photon pair should match the memory frequency and telecom-band, respectively. In the situation of frequency mismatching, frequency conversion may be required \cite{kumar1990quantum, huang1992observation}.

\section{Mathematical derivation}
\label{sec:math}

\subsection{General framework of rates calculation}
\label{subsec:gen_frame}

\subsubsection{Two-link situation}
\label{subsubsec:two_link_situ}
Let us start with the entanglement generation process, which is probabilistic. One has to try many times until the entanglement is successfully generated. After each attempt, one has to wait for the Bell-state measurement signal that tells whether the attempt is successful. If not, the memories need to be emptied and one needs to try again. In this paper, we do not consider the time of memory reinitialization (negligible compared with communication time) and thus the time required for each attempt is $T_{0} = L_{0}/c$, where $L_{0}$ is the length of elementary link and $c=2*10^{8}$ $\mathrm{ms^{-1}}$ is the speed of light in optical fiber. Considering a entanglement generation probability of $p_{0}$ and the case that entanglement is generated until $n$th attempt, the probability distribution function(PDF) for $n$ is thus
\begin{equation}
    P(n) = p_{0}(1 - p_{0})^{n - 1}.
\label{equ:basic_pdf}
\end{equation}
We do not consider the dark count of the detector since it can be considerable small. Superconducting nanowire single-photon detectors with 30Hz dark count rate are already used in measurement-device-independent quantum key distribution \cite{yin2016measurement} and with milli-Hz dark count rate is also demonstrated in lab \cite{schuck2013waveguide}.

Now let us consider the repeater with two links, where the entanglement is generated independently with probability $p_{0}$ for each link. The entanglement swapping can be performed only after the entanglement is established in both links. We define variables $n_{1}$ and $n_{2}$ as the number of attempts to establish the two links, respectively, and thus the joint PDF for these two variables is $P(n_{1}, n_{2}) = p_{0}^{2}(1 - p_{0})^{n_{1} + n_{2} - 2}$. We further define three variables $n_{max}$, $n_{min}$, and $n_{dif}$, denoting $\mathrm{max}\{n_{1}, n_{2}\}$, $\mathrm{min}\{n_{1}, n_{2}\}$ and $|n_{1} - n_{2}|$, respectively. Obviously, they are related by $n_{max} = n_{min} + n_{dif}$. It is useful to show the probability distribution function of $n_{dif}$ 
\begin{equation}
\label{equ:prob_dif}
     p(n_{dif}) = \left\{
                            \begin{aligned}
                            & \frac{2p_{0}(1-p_{0})^{n_{dif}}}{2-p_{0}} & n_{dif} \neq 0 \\
                            & \frac{p_{0}}{2 - p_{0}}& n_{dif} = 0 
                            \end{aligned}
                  \right.
\end{equation}
and the expectation of these variables
\begin{equation}
\begin{split}
\label{equ:exp_mmd}
    \langle n_{max}\rangle & = \frac{3 - 2p_{0}}{(2 - p_{0})p_{0}};  \langle n_{min}\rangle =  \frac{1}{(2 - p_{0})p_{0}}\\
    \langle n_{dif}\rangle & = \frac{2 - 2p_{0}}{(2 - p_{0})p_{0}}.
\end{split}
\end{equation}
We call $n_{max}T_{0}$ the ``preparation time", and $n_{dif}T_{0}$ the ``decay time"(waiting time). This is to say, the entanglement swapping is processed after the preparation time, and the entangled state of one link is destroyed during the decay time. The memory decay, as it will not decrease the fidelity but the efficiency \cite{staudt2007interference}, is modeled as the following general situation
\begin{equation}
\begin{split}
    \label{equ:decay}
    & \alpha\ket{\Psi}\bra{\Psi} + (1 - \alpha) \rho \xrightarrow{decay} \\
    & e^{-\Delta t/\tau_{M}} \alpha\ket{\Psi}\bra{\Psi} + (1 - e^{-\Delta t/\tau_{M}} \alpha)\rho',
\end{split}
\end{equation}
where $\ket{\Psi}\bra{\Psi}$ represents the maximized entangled state that we are interested in, $\rho$ and $\rho'$ are `unwanted states' that have no contribution to the repeater performance, and $\Delta t$ and $\tau_{M}$ are the decay time and the lifetime of the memory, respectively. It is important to notice that the decay time is the waiting time for the single-photon BSM, while it is two times waiting time for two-photon BSM since both memories in a link will decay. The entanglement swapping probability $p_{s}$ and resultant state, therefore, depend on the decay time, and thus $n_{dif}$. This dependence can be understood via the following calculation: before swapping, one has to establish two neighboring links, which could not be perfect and thus we consider two mixed states
    \begin{equation}
    \label{equ:mat_before}
        \begin{split}
            \rho_{1} & = \alpha_{1} \ket{\Psi_{1}} \bra{\Psi_{1}} + (1 - \alpha_{1})\rho_{1}\\
            \rho_{2} & = \alpha_{2} \ket{\Psi_{2}} \bra{\Psi_{2}} + (1 - \alpha_{2})\rho_{2},
        \end{split}
    \end{equation}
    where $\ket{\Psi_{1}}$ and $\ket{\Psi_{2}}$ are the entangled state and $\rho_{1}$, $\rho_{2}$ are the unwanted state. After swapping, the new state can be expressed as 
    \begin{equation}
        \rho  = \alpha \ket{\Psi_{12}} \bra{\Psi_{12}} + (1 - \alpha)\rho\\,
    \end{equation}
where $\ket{\Psi_{12}}$ is the entangled state determined by the Bell-state measurement, and $\rho$ is the unwanted state. It is important to notice that the unwanted state in the single-photon BSM is a pure vacuum state, while in two-photon BSM it contains the resultant states from both single-memory decay and two-memory decay(vacuum state). For the two-photon BSM, the swapping will eliminate all other cases than the entangled state. The expression of $\alpha$ and the success probability $p_{s}$, depend on the type of Bell-state measurement(BSM). Single-photon BSM will create a vacuum state since it cannot exclude the situation where the two photons are stored in the two local nodes in neighboring links. Therefore, the fidelity of resultant state $\alpha$ is 
\begin{equation}
\label{equ:alpha_after}
        \alpha = \frac{\alpha_{1}\alpha_{2}}{\alpha_{1} + \alpha_{2} - \alpha_{1}\alpha_{2} \eta_{d}},
\end{equation}
and the success probability is
\begin{equation}
\label{equ:prob_after}
    p_{s} = \frac{1}{2}(\alpha_{1}\eta_{d} + \alpha_{2}\eta_{d} - \alpha_{1}\alpha_{2}\eta_{d}^{2}),
\end{equation}
where $\eta_{d}$ is the detector efficiency. It is important to note that the definition of fidelity and success probability is different from the definition in Ref. \cite{Sangouard}, where we substitute the product of detector efficiency and memory efficiency $\eta_{d}\eta_{m}$ with only detector efficiency $\eta_{d}$. This is because
we will consider the memory efficiency in the entanglement generation process, where an unsuccessful storage or retrieval of the photon will create a vacuum state, and thus $\alpha_{1}$ and $\alpha_{2}$ depend on the memory efficiency. 

On the other hand, the resultant state of two-photon BSM should be a pure entangled state and thus the fidelity is
\begin{equation}
    \alpha = 1.
\end{equation}
The success probability is simply
\begin{equation}
    \label{equ:prob_after2}
        p_{s} = \alpha_{1}\alpha_{2}\eta_{d}^{2} / 2,
\end{equation}
where the half is the intrinsic success probability of usual two-photon BSM.

In the entanglement swapping process, if the first link establishes the entanglement first, the memory in this link would decay and thus $\alpha_{1} = \alpha_{0}\mathrm{exp}(-n_{dif}T_{0}/\tau_{M})$, $\alpha_{2} = \alpha_{0}$, where $\alpha_{0}$ is the entangled state fidelity after entanglement generation. Therefore, the swapping probability in both case, and the state fidelity after swapping in the single-photon BSM situation depend on $n_{dif}$. 

Now let us calculate the average entanglement distribution time(EDT) for the two-link situation. Without loss of generality, we consider a successful entanglement swapping after $r$th swapping attempts and the EDT for this case is given by the following expression:
\begin{equation}
    \label{equ:time1}
   T_{tot}^{(1)} = (\sum_{k = 1}^{r} n_{k, max}T_{0}) p_{r, s}\prod_{k' = 1}^{r-1}(1 - p_{k', s}),
\end{equation}
where the subscript ``$k$" denotes the $k$th swapping attempt, and therefore $\sum_{k = 1}^{r} n_{k, max}T_{0}$ is the total preparation time in total $r$ swapping attempts and $p_{r, s}\prod_{k' = 1}^{r-1}(1 - p_{k', s})$ is the corresponding probability. The expectation of the EDT also requires averaging $r$,  $n_{k, max}$, and $n_{k, dif}$, which is difficult since both $p_{k, s}$ and $n_{k, max}$ depends on $n_{k, dif}$. It is easier to rewrite $n_{k, max}$ as $n_{k, dif} + n_{k, min}$, where $n_{k, min}$ is independent with $n_{k, dif}$. It is however, still difficult to calculate analytically the result without assumptions on $\langle n_{k, dif} * p_{k, s} \rangle$, where $p_{k, s}$ depends on the waiting time and therefore with $n_{k, dif}$. We define a new variable $\beta = \langle n_{k, dif} * p_{k, s} \rangle / \langle n_{k, dif} \rangle \langle p_{k, s} \rangle$, and clearly, we have
\begin{equation}
\label{equ:beta_range}
    0 < \beta < 1,
\end{equation}
where the right side `$<$' is because $n_{k, dif}$ and $p_{k, s}$ are negatively correlated. The numerical evidence discussed in Appendix \ref{app:num_evi} suggest that in single-photon BSM,
\begin{equation}
\label{equ:beta_sps}
    \beta \approx 1.
\end{equation}
On the other hand, in the two-photon BSM, $\beta$ can reach lower bound and upper bound in Eq. (\ref{equ:beta_range}) in different regimes: in the low-$p_{0}$ regime, $\beta \approx 1$, while in the high-$p_{0}$, high-lifetime regime, $\beta \approx 0$.


With Eq. (\ref{equ:beta_sps}) and Eq. (\ref{equ:beta_range}), we can simplify Eq. (\ref{equ:time1})
\begin{equation}
    \label{equ:time11}
  \langle T_{1tot}^{(1)} \rangle = \frac{1}{\langle p_{s} \rangle}(\langle n_{dif} \rangle + \langle n_{min} \rangle)T_{0} = \frac{\langle n_{max} \rangle}{\langle p_{s} \rangle}T_{0},
\end{equation}
for single-photon BSM and 
\begin{equation}
    \label{equ:time12}
 \frac{\langle n_{min} \rangle}{\langle p_{s} \rangle}T_{0} < \langle T_{2tot}^{(1)} \rangle < \frac{\langle n_{max} \rangle}{\langle p_{s} \rangle}T_{0}.
\end{equation}
for two-photon BSM, where the detailed derivation can also be found in Appendix \ref{app:proof}. It is important to notice that for two-photon BSM, $T_{2tot}^{(2)}$ can be estimated using the average of the lower and upper bound $({\langle n_{min} \rangle}T_{0}/ {\langle p_{s} \rangle} + {\langle n_{max}\rangle}T_{0}/ {\langle p_{s} \rangle}) /2 = T_{0}/p_{0}\langle p_{s}\rangle$. As we can see from Eq. (\ref{equ:exp_mmd}), in the worst case $\langle n_{max} \rangle$ is no more than $3\langle n_{min} \rangle$, and the error rate of this estimation is thus no more than $50\%$, which is still a good estimation since repeater rates vary over many orders of magnitude.

\subsubsection{Postselection and beyond two links}
Though we have given the EDT for repeater schemes with single-photon BSM in Eq. (\ref{equ:time11}), the entanglement is imperfect and cannot be used directly for quantum communication purpose and thus postselection is necessary. To implement postselection, a separate chain of two links is placed in such a way that the two end nodes are placed at the same location as the two end nodes from the original chain, respectively. After entanglement is established in both chain, single-photon BSM is performed at each end and projected the state into a two-photon entanglement state \cite{duan2001long}. The treatment of the postselection is analog to the four-link situation(nesting level of 2). It is important to note EDT beyond two links is interesting in general, even if our motivation here is to include postselection. 

We consider the establish time $T_{1}$ and $T_{2}$, and state fidelity $\alpha_{1}$ and $\alpha_{2}$, respectively for two sublinks. The average establish time and the state fidelity is given in Eq. (\ref{equ:time11}) and Eq. (\ref{equ:alpha_after}). Here we follow the same procedure in the two-link situation and consider the postselection is successful at the $j$th attempt. The EDT is expressed as 
\begin{equation}
\label{equ:time_post}
    T^{(2)}_{tot} = (\sum_{i = 1}^{j} T_{i, max})p_{j, ps}\prod_{i' = 1}^{j - 1}(1 - p_{i', ps}),
\end{equation}
where the subscript ``$i$" represent the $i$th postselection attempt, $p_{i, ps} = \alpha_{i, 1} \alpha_{i, 2} \mathrm{exp}(-T_{i, dif}r)$ is the postselection probability for $i$th attempt, $T_{i, max} = \mathrm{max}\{T_{i, 1}, T_{i, 2}\}$, and $T_{i, dif} = |T_{i, 1} - T_{i, 2}|$. Hence, $\sum_{i = 1}^{j} T_{i, max}$ is the total time for a success in $j$th postselection attempt and $p_{j, ps}\prod_{i' = 1}^{j - 1}(1 - p_{i', ps})$ is the corresponding probability. The calculation also requires averaging $j$, $T_{i, max}$ and $T_{i, dif}$, which is difficult since both $T_{i, max}$ and $p_{i, ps}$ depends on them and we do not know the exact probability distribution function. Fortunately, we can bypass the problem with solid approximations. First, it is safe to claim that $\alpha_{i, 1}$ is independent with $T_{i, 1}$ and same for $\alpha_{i, 2}$. This is because $\alpha_{i, 1}$ is defined in Eq. (\ref{equ:alpha_after}) and it only depends on the waiting time in the first entanglement swapping. Thus, we rewrite the postselection probability as
\begin{equation}
\label{equ:ass_post}
    p_{i, ps} = \langle \alpha_{i, 1}\rangle \langle \alpha_{i, 2} \rangle \mathrm{exp}(-T_{i, dif}r).
\end{equation}
Then, we substitute $T_{i, max}$ with $T_{i, min} + T_{i, dif}$, where $T_{i, min} = \mathrm{min}\{T_{i, 1}, T_{i, 2}\}$ and is independent with $T_{i, dif}$. If we define $\beta' = \langle T_{i, dif} p_{i, ps}\rangle / \langle T_{i, dif} \rangle \langle p_{i, ps}\rangle$, similar to Eq.\ref{equ:beta_range}, we have
\begin{equation}
\label{equ:ass_t_dif}
    0 < \beta' < 1,
\end{equation}
which gives the lower bound and upper bound of $T_{tot}^{(2)}$
\begin{equation}
\label{equ:total_time_post}
    \frac{\langle T_{min}\rangle}{\langle \alpha \rangle^{2} \langle \mathrm{exp}(-T_{dif} r) \rangle} < \langle T^{(2)}_{tot} \rangle < \frac{\langle T_{max}\rangle}{\langle \alpha \rangle^{2} \langle \mathrm{exp}(-T_{dif} r) \rangle},
\end{equation}
where we have used $\langle \alpha_{i, 1} \rangle = \langle \alpha_{i, 2} \rangle = \langle \alpha \rangle$ So far, we have not made assumptions on the probability distribution function on $T_{1}$ and $T_{2}$, which is necessary to derive the expectation value of $T_{min}$, $T_{max}$, and $\mathrm{exp}(-T_{dif} r)$. Here we assume the establish time for one sublinks, i.e., two links, is $mT_{0}$, where the probability distribution function of $m$ is $P(m)$ that defined in Eq. (\ref{equ:basic_pdf}) except substituting $p_{0}$ for 
\begin{equation}
\label{equ:ass_prob_post}
    p_{0}' = 2p_{0}\langle p_{s} \rangle/3.
\end{equation}
This assumption gives the same expectation value of establish time for two links as given in Eq. (\ref{equ:time11}), while the probability distribution function is different. In fact, the numerical evidence in Appendix \ref{app:num_evi} shows that by substituting $p_{0}$ for $p_{0}'$ in Eq. (\ref{equ:prob_dif}) and Eq. (\ref{equ:exp_mmd}), it gives a good approximation of the probability distribution function of $T_{dif}$ and expectation of $T_{min}$ and $T_{max}$. Thus, we can calculate the lower bound and the upper bound in Eq. (\ref{equ:total_time_post}), and similar to the treatment in the two-photon BSM situation, we use the average of the lower bound and the upper bound to approximate $\langle T^{(2)}_{tot} \rangle$
\begin{equation}
\label{equ:total_time_post_sps}
     \langle T^{(2)}_{tot} \rangle \approx \frac{\langle T_{max}\rangle + \langle T_{min}\rangle}{2\langle \alpha \rangle^{2} \langle \mathrm{exp}(-T_{dif} r) \rangle}.
\end{equation}

\subsection{Two-link repeater performance for different schemes}
One can calculate the average entanglement distribution time from Eq. (\ref{equ:total_time_post_sps}) and Eq. (\ref{equ:time12}) for single-photon BSM and two-photon BSM. respectively. Here we give a detailed procedure of calculating repeater rates for each repeater scheme.

\subsubsection{The ``1 + 1" scheme}
In the ``1 + 1" scheme, the single photon emitted from each source will be partially transmitted to the central beam-splitter and the memory (controlled by a local beam-splitter). With one click after the central beam-splitter, i.e., the single-photon Bell-state measurement, the entanglement is claimed to be a success while a mixed state $\alpha^{(0)} | \Psi \rangle\langle\Psi|+(1-\alpha^{(0)} )| 0\rangle\langle 0|$ is produced. The success probability is $2 \gamma(1 - \gamma) \eta_{t} \eta_{s} \eta_{d} + 2\gamma^{2} \eta_{t}(1 - \eta_{t}) \eta_{s} \eta_{d}$, where $\gamma$ is the transmission coefficient, $\eta_{t} = \mathrm{exp}(-L_{0}/L_{att})$ is the transmission loss, $L_{att}$ the fiber attenuation length, and $\eta_{s}$ is the single-photon source efficiency. The first term is the case that only one photon is sent to the central beam-splitter, while the second term represents both photons being sent to the central beam-splitter while one of the photons is lost due to the fiber attenuation. The success probability of entanglement generation can be approximated as 
\begin{equation}
    p_{0} = 2 \gamma \eta_{t}\eta_{s}\eta_{d},
\end{equation}
since $\eta_{t} \ll 1$. The fidelity of resultant state with unity-efficiency quantum memory is $1 - \gamma$, while an unsuccessful storage or retrieval of the photon will cause a vacuum component. Thus, considering memory efficiency $\eta_{m}$, the fidelity of the resultant state is
\begin{equation}
    \alpha^{(0)} = \eta_{m}(1 - \gamma).
\end{equation}
It is important to note that memory efficiency here is the product of storage efficiency and the retrieve efficiency, i.e., efficiency without considering decay.

Based on Eq. (\ref{equ:alpha_after}) and Eq. (\ref{equ:prob_after}) and with $\alpha_{1} = \alpha^{(0)}$ and $\alpha_{2} = \alpha^{(0)} \mathrm{exp}(-n_{dif}^{(1)} r)$, we can derive the average fidelity after swapping
\begin{equation}
    \langle \alpha^{(1)} \rangle = \langle \frac{\alpha^{(0)}\mathrm{exp}(-n_{dif}^{(1)} r)}{1 + (1 - \alpha^{(0)}\eta_{d})\mathrm{exp}(-n_{dif}^{(1)} r)} \rangle,
\end{equation}
and the average swapping probability
\begin{equation}
    \langle p_{s}^{(1)} \rangle = \frac{\alpha^{(0)}\eta_{d}}{2} \langle 1 + (1 - \alpha^{(0)} \eta_{d}) \mathrm{exp}(-n_{dif}^{(1)} r) \rangle.
\end{equation}
The average entanglement distribution time can now be calculated via Eq. (\ref{equ:total_time_post_sps}).

\subsubsection{The ` $\tilde{2}$ + 1' scheme}
The ``$\tilde{2}$ + 1'' scheme is similar to the DLCZ scheme since the probability of photon-pair emission should be small to suppress multi-pair emission. It is important to notice that in Sec. \ref{subsec:gen_frame} we have defined source efficiency $\eta_{s}$, which refers to the probability to extract a photon in deterministic photon source, while the analog in non-deterministic photon source is the emission probability. In this paper, we do not distinguish these two terms and call them ``source efficiency'' $\eta_{s}$. The success probability of the state after entanglement generation can be easily derived as 
\begin{equation}
    p_{0} = 2\eta_{t}\eta_{s}\eta_{d}
\end{equation}
As $\eta_{s} \ll 1$, one can ignore the situation of coincident emission, and thus the fidelity of resultant state is 
\begin{equation}
    \alpha^{(0)} = \eta_{m}.
\end{equation}
Similarly, the average fidelity after swapping is 
\begin{equation}
\langle \alpha^{(1)} \rangle = \langle \frac{\mathrm{exp}(-n_{dif}^{(1)} r)}{1 + (1 - \eta_{d})\mathrm{exp}(-n_{dif}^{(1)} r)} \rangle,
\end{equation}
and the average swapping probability is
\begin{equation}
    \langle p_{s}^{(1)} \rangle = \frac{\eta_{d}}{2} \langle 1 + (1 -  \eta_{d}) \mathrm{exp}(-n_{dif}^{(1)} r) \rangle.
\end{equation}

\subsubsection{The ``2 + 2" scheme and the `` $\tilde{2}$ + 2" scheme}
The calculation for the ``2 + 2" scheme and the ``$\tilde{2}$ + 2" scheme are the same, but it is worth noting that $\eta_{s}$ in the non-deterministic source is much smaller than the deterministic source. The probability generation probability for the ``2 + 2" scheme and ``$\tilde{2}$ + 2" scheme is
\begin{equation}
    p_{0} = \eta_{t}^{2} \eta_{s}^{2} \eta_{d}^{2} / 2,
\end{equation}
where the one half is the intrinsic success probability of two-photon BSM. The fidelity of the created mixed state is
\begin{equation}
    \alpha^{(0)} = \eta_{m}^{2}.
\end{equation}
The average success probability can thus be derived from Eq. (\ref{equ:prob_after2})
\begin{equation}
    \label{equ:ps_2+2}
    \langle p_{s} \rangle = \eta_{d}^{2} \eta_{m}^{4} \langle \mathrm{exp}(-n_{dif}r) \rangle /2.
\end{equation}

\section{Implementation}
\label{sec:imp}

In this section, we consider physical platforms for both quantum memories and photon sources. For quantum memories, we focus on rare-earth-ion based memory \cite{siyushev2014coherent, xia2015all, fraval2005dynamic, longdell2005stopped, Zhong2015, Rancic2017, sabooni2010storage,usmani2010mapping,AMARI20101579,Bonarota_2011, timoney2012atomic, Jobez_2014,sabooni2013efficient} and Rydberg atom-ensemble memory \cite{PhysRevLett.87.037901, saffman2002creating, PhysRevA.97.053803, PhysRevLett.123.140504, yang2016efficient}, which are evaluated in terms of memory lifetime and efficiency. As for photon sources, we consider quantum dots as SPS \cite{senellart2017high, ding2016demand} and dPPS \cite{wang2019demand}, Rydberg atoms as SPS \cite{dudin887, dudin2012observation} and semi-dPPS \cite{li2019semi}, and spontaneous parametric down-conversion source as ndPPS \cite{caspani2017integrated, PhysRevLett.121.250505}. Thus, we propose the implementations for repeater schemes we mentioned in Sec. \ref{sec:reps}, shown in Tab. \ref{tab:rep_imp}.

\begin{table}[ht]
    \caption{\label{tab:rep_imp}Repeater scheme implementations}
    \begin{ruledtabular}
    \begin{tabular}{ c | m{5cm} } 
    Schemes & Implementations\\
    \hline
    $1 + 1$ & QDs + REIs; RAs\\
    \hline
    $2 + 2$ & QDs + REIs; RAs  \\
    \hline
    $\tilde{2} + 1$ & PDC + REIs \\
    \hline
    $\tilde{2} + 2$ & PDC + REIs\\
    \end{tabular}
    \end{ruledtabular}
\end{table}

\subsection{Rare-earth-ions(REIs) based quantum memory}
Rare-earth-ion doped crystals are attractive as quantum memories \cite{zhong2019emerging}, especially on storage efficiency \cite{hedges2010efficient, sabooni2013efficient}, multimode capacity \cite{usmani2010mapping, bonarota2010efficiency}, and polarization qubit storage \cite{zhou2012realization, gundougan2012quantum}. At cryogenic temperatures, rare-earth-ion doped crystals exhibit long ground state coherence time: the electron spin coherence time of milliseconds is seen in many experiments \cite{siyushev2014coherent, xia2015all}, and the nuclear spin coherence time can reach seconds or even hours \cite{fraval2005dynamic, longdell2005stopped, Zhong2015, Rancic2017}. 

The relevant transitions in rare-earth ions in solids have narrow homogeneous lines in combination with large inhomogeneous broadening. This can be used to create a periodic structure of narrow absorption peaks, so-called atomic frequency combs(AFC) \cite{PhysRevA.79.052329}. AFC is of great interest in the repeater applications since it allows efficient storage and readout of multiple temporal modes, which could greatly enhance the repeater performance \cite{collins2007multiplexed}.  The temporal modes range from dozens to thousands with a typical storage time of milliseconds level and efficiencies in free space range from 1-35\% \cite{sabooni2010storage,usmani2010mapping,AMARI20101579,Bonarota_2011, timoney2012atomic}. Though theoretically upper bound for AFC quantum memory efficiency is unity, the requirement of a large optical depth \cite{PhysRevA.79.052329} is hard to achieve. Putting the memory inside an asymmetric optical cavity can greatly enhance the efficiency by meeting ``impedance matching" condition \cite{afzelius2010impedance}. So far, 53\% \cite{Jobez_2014} and 56\% \cite{sabooni2013efficient} memory efficiency have been reported in cavity-based AFC memory, with milliseconds storage time. 

\subsection{Rydberg atoms (RAs): Rydberg-state based photon sources and ground-state quantum memory}
Rydberg states are characterized by a high principal quantum number and a corresponding large size \cite{saffman2010quantum}. Due to the large dipole moments and strong dipole-dipole interactions, the excitation of Rydberg atoms would shift the excited energy levels of nearby atoms, which excludes the resonance excitation of these atoms and is called the Rydberg-blockade \cite{PhysRevLett.87.037901}. The Rydberg-blockade, as the central topic of Rydberg atoms' application in quantum information processing, makes it possible to realize photon sources \cite{dudin887, dudin2012observation, li2019semi} and quantum memories \cite{PhysRevLett.87.037901, saffman2002creating, PhysRevA.97.053803, PhysRevLett.123.140504, yang2016efficient}.

Both SPS and semi-dPPS can be realized in Rydberg atoms ensembles. The SPS relies on the Rydberg blockade effect, where the shifting in the energy level will inhibit transition into all but single excitation state \cite{PhysRevLett.87.037901}. The spin-wave state is then converted into a light field by retrieving the excitation back to the intermediate level \cite{dudin887}. Room temperature SPS has also been demonstrated, though with low efficiency(4\%) \cite{dudin2012observation}. The semi-dPPS can also be realized via the Rydberg blockade effect, where two excitations with different momentum can be entangled. After mapping one excitation to a photon, a photon and an atomic excitation can be entangled in the polarization domain. With further retrieving the ground-state excitation, the atom-photon entanglement is converted into a photon pair entanglement \cite{li2019semi}. The intrinsic efficiency of this method is $50\%$, which is semi-deterministic. In the future, with more efficient qubits manipulation, $100\%$ intrinsic efficiency, i.e., nPPS, can also be realized.

Cold atoms are also attractive as quantum memories since they allow both rapid and deterministic preparation of quantum states and their efficient transfer into single-photon light fields \cite{PhysRevLett.87.037901, saffman2002creating}. Although the optical lifetime of highly excited Rydberg atoms can reach several hundred microseconds \cite{saffman2010quantum}, the optical coherence time is only several microseconds \cite{PhysRevA.97.053803} because Rydberg atoms are sensitive to the environment. Thus, to realize a long memory lifetime, we need to transfer the Rydberg excitation to a long-lived ground state. The demonstrated mapping efficiency from the Rydberg state to the ground state is already more than $70\%$ \cite{PhysRevLett.123.140504}. The coherence time of the excitation stored in the ground state, or spin-wave is mainly limited by the motion of the atoms and the fluctuation of the residual magnetic field. Combing optical lattice,``clock state" storage and cavity enhancement read-out, 220ms spin-wave lifetime and initial intrinsic retrieval efficiency of $76\%$ have been demonstrated \cite{yang2016efficient}. It is worth noting that the size of the Rydberg blockade radius poses a limitation on the number of atoms that can be used. The low optical depth will decrease the coupling strength between the single-photon and the ensemble, and thus the retrieval efficiency. It is therefore necessary to couple the ensemble with a cavity to enhance the overall efficiency.

\subsection{Quantum dots(QDs) based single-photon and photon-pair source}

Quantum dots(QDs) \cite{lodahl2015interfacing} are recognized as one of the best on-demand single-photon sources that possess the highest quantum efficiency in solid-state quantum emitter schemes \cite{senellart2017high}.  In one experiment, near-perfect single-photon purity(99.1\%), indistinguishability(98.6\%), and high extraction efficiency(66\%) have been reported based on resonant excitation of InAs-GaAs QDs in a micropillar cavity \cite{ding2016demand}. 

Photon-pair sources can also be realized by radiative cascades in quantum dots. 
 In a recent experiment, high fidelity(90\%), pair extraction efficiency(62\%), and indistinguishability(90\%) are demonstrated by a single InGaAs quantum dot coupled to a circular Bragg grating bullseye cavity with broadband high Purcell factor up to 11.3 \cite{wang2019demand}.

It is important to notice that the overall efficiency is usually limited by the scattering loss and fiber coupling efficiency, and a typical value of overall efficiency is only $15\%$ \cite{ding2016demand}. Fortunately, the photonic nanowire approach to fabricating efficient quantum light sources has been proposed \cite{friedler2009solid} and shown great promise to achieve high extraction efficiency and high fiber coupling efficiency. The collective efficiency has achieved $72\%$ in single-photon source based on InAs QDs embedded in a GaAs photonic nanowire \cite{claudon2010highly}. The photon-pair source has also been reported in nanowire quantum dots \cite{versteegh2014observation, huber2014polarization, jons2017bright} with extraction efficiency around 15$\%$. With further optimization of the nanowire shape, the extraction efficiency of more than 90$\%$ can be expected \cite{friedler2009solid}.

\subsection{Parametric down-conversion(PDC) based photon-pair source}
One of the most widely-used techniques to produce entangled photon-pair is by spontaneous nonlinear parametric processes. The process that one photon in the pumping laser goes through materials with second-order(\(\chi^{(2)}\)) can annihilate into two photons is spontaneous parametric down-conversion(SPDC). A similar process for third-order(\(\chi^{(3)}\)) materials is called spontaneous four-wave mixing(SFWM) \cite{caspani2017integrated}. Photons can be entangled in polarization, frequency, and time. A recent outstanding polarization-entanglement source uses narrow-band spectral filters that eliminate spectral correlations and has demonstrated high indistinguishability(97\%) and purity(99\%) \cite{PhysRevLett.121.250505}. It is important to note that, in order to suppress the multi-pair emission, the emission probability should be low and thus the photon-pair generation process is non-deterministic. On the other hand, it is relatively easy for this type of source to match the emission frequencies of each photon to a desired wavelength. For example, one photon can be in resonance with a quantum memory, while the other one matches the telecom band of optical fibers. 

\begin{figure*}
\centering
    \includegraphics[width=\linewidth, height=0.6\linewidth]{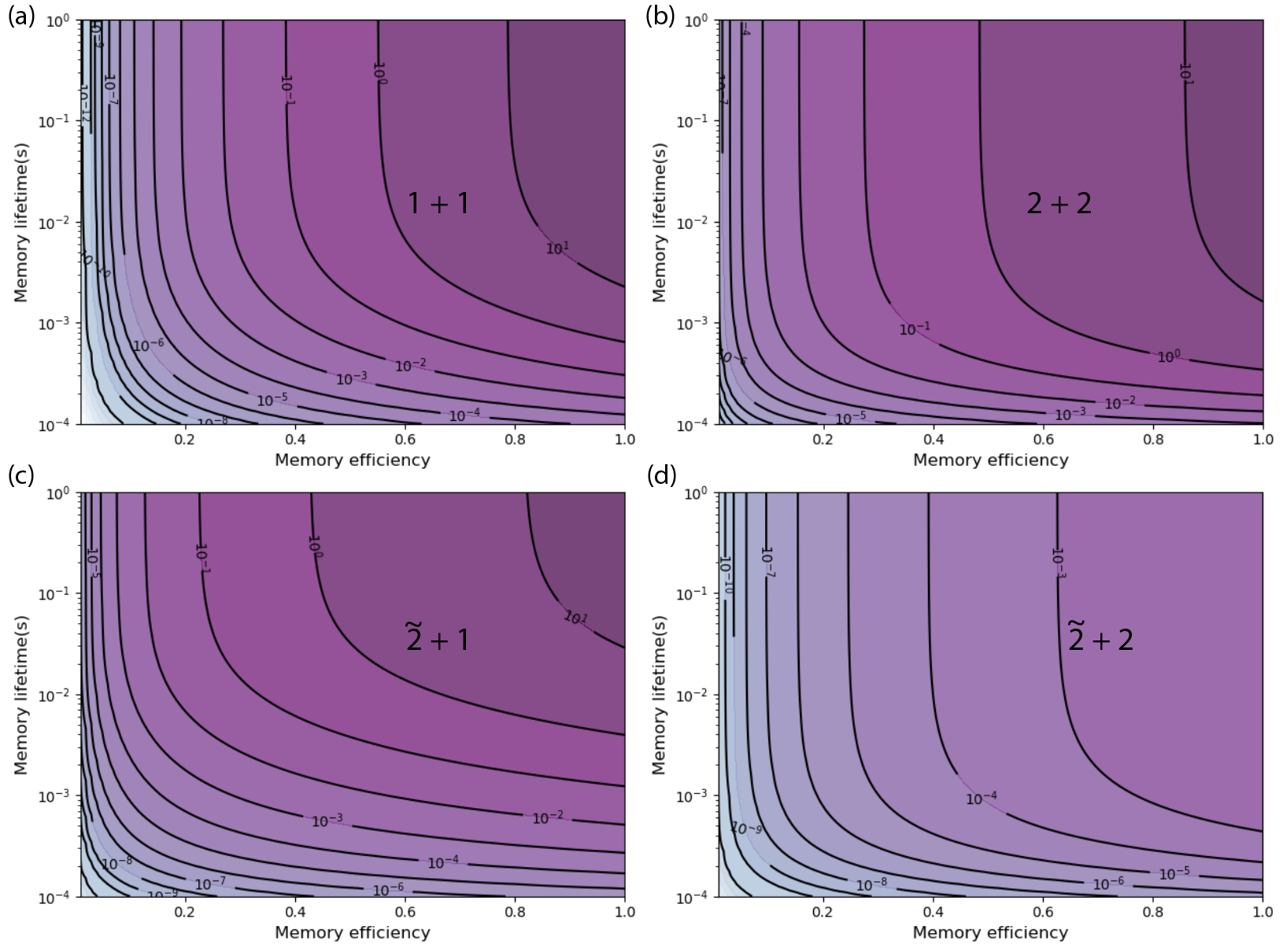}
    \caption{(color online.) Rates of various repeater schemes for a total distance of 100km. The numbers in the contour line represent the corresponding repeater rates in Hz. The plot shows the situation of two links(nesting level is 1). The corresponding repeater protocols and parameter regimes are (a) SPS + single-photon BSM($1 + 1$) with local beam-splitter transmission probability $0.8$ and single-photon emission probability $0.75$; (b)dPPS + two-photon BSM($2 + 2$) with photon-pair emission probability $0.5$; (c)ndPPS + single-photon BSM($\tilde{2} + 1$) with photon-pair emission probability $0.03$; (d) ndPPS + two-photon BSM($\tilde{2} + 2$) with photon-pair emission probability $0.03$.}
    \label{fig:eff_vs_time}
\end{figure*}


\section{Numerical results}\label{sec:numer}

\subsection{Memory requirements}
We first study the repeater rates as a function of memory lifetime and efficiency. It is worth noting that the efficiency of photon sources should be predetermined to use Eq. (\ref{equ:total_time_post}) and Eq. (\ref{equ:time12}) to calculate the repeater rates, defined as the reciprocal of EDT. We consider high, but realistic efficiencies of $75\%$ and $50\%$ for SPS and PPS, respectively, and the photon-pair emission probability $3\%$ for ndPPS. The transmission coefficient $\gamma=0.2$ in the ``1 + 1" scheme, the fiber attenuation length $L_{att}=22$km (for telecom-wavelength range around 1550nm), and the detector efficiency $\eta_{d} = 0.95$. In Fig. \ref{fig:eff_vs_time} we show the repeater rates for a total distance of 100 km, where the contour line represents the same repeater rates for various parameters of memory lifetime and efficiency. 

The figures can provide useful information. To begin with, the graphs show the potential trade-off between memory lifetime and efficiency to achieve a target repeater rate, say 1 Hz. For example, in the ``2 + 2'' scheme, to realize the target repeater rate, one can use memories with $1$ms lifetime and $50\%$ efficiency, or with unity efficiency and $0.2$ms lifetime, or with $15\%$ efficiency and $1$s lifetime. 

One can find the most efficient way to improve the repeater performance by improving the memory parameter along the gradient in the contour graphs. In particular, one can see that for a short lifetime but high efficiency, there is limited benefit in improving the efficiency further and vice versa. Conversely, in the high lifetime regime (e.g. $>$ 10 ms for the ``2 + 2" scheme), the gradients of the contour lines are parallel to the efficiency axis, meaning that the increase in efficiency will considerably improve the repeater rates. We also notice that the contour line is more concentrated in the small memory lifetime($<$ 1ms) and efficiency($<$ 20\%) regime, which means an improvement in lifetime or efficiency respectively in the corresponding regimes will dramatically improve the repeater rates.

Moreover, the graphs give the upper bound of the repeater rates that can be achieved in these schemes for this distance. The maximum repeater rate under the present assumptions is of order 10 Hz in the ``1 + 1", ``2 + 2", and ``$\tilde{2} + 1$" scheme and $10^{-3}$ Hz in the ``$\tilde{2} + 2$" scheme. These upper bounds of the repeater rates, which correspond to perfect quantum memories, can be improved by using better sources, more links, or multiplexing \cite{Sangouard}. 

These results were obtained for exponential decay of the memory efficiency as described in section III. In Appendix \ref{app:com} we compare our results to what one would obtain under the common simplified assumption of a memory cut-off time. The main conclusion is that the cut-off is not a good approximation for short lifetimes.  

\subsection{Comparison of implementations}
Let us now consider the practical implementations in Tab. \ref{tab:rep_imp}, and give the expected repeater rates with realistic parameter regimes. 

The ``1 + 1'' scheme can be implemented using QDs and REIs, or RAs, where in the latter case the RA ensemble can serve as both the single-photon source and the memory, with the beam splitter operation being completed by partial readout. The parameters we will use in the numerical calculation is: transmission probability of local beam-splitter is 0.2; efficiency of QDs based SPS $\eta_{s(QDs)} = 75\%$; efficiency and lifetime of REIs based memory $\eta_{m(REIs)} = 70\%$, $\tau_{m(REIs)} = 1\mbox{ms}$; efficiency of RAs based source $\eta_{s(RAs)} = 15\%$; efficiency and lifetime of RAs based memory $\eta_{m(RAs)} = 75\%$, $\tau_{m(RAs)} = 220\mbox{ms}$.

For the ``2 + 2'' scheme, we consider the same platforms as in the the ``1 + 1'' scheme-QDs and REIs, or RAs, except that the photon sources are (semi-)dPPS. The parameters we will use in the numerical calculation is: efficiency of QDs based SPS $\eta_{s(QDs)} = 50\%$; efficiency and lifetime of REIs based memory $\eta_{m(REIs)} = 70\%$, $\tau_{m(REIs)} = 1\mbox{ms}$; efficiency of RAs based source $\eta_{s(RAs)} = 15\%$; efficiency and lifetime of RAs based memory $\eta_{m(RAs)} = 75\%$, $\tau_{m(RAs)} = 220\mbox{ms}$.

For the ``$\tilde{2}$ + 1'' and the ``$\tilde{2}$ + 2'' scheme, we propose to use PDC and REIs. It is important to note the emission probability should be low to suppress the multiphoton error. The parameters we will use in the numerical calculation is thus: probability of single-pair emission for PDC source $\eta_{s(PDC)} = 3\%$; efficiency of rare-earth-ions based memory $\eta_{m(REIs)} = 70\%$; lifetime of rare-earth-ions based memory $\tau_{m(REIs)} = 1\mbox{ms}$.

\begin{figure}
    \centering
    \includegraphics[width=\linewidth]{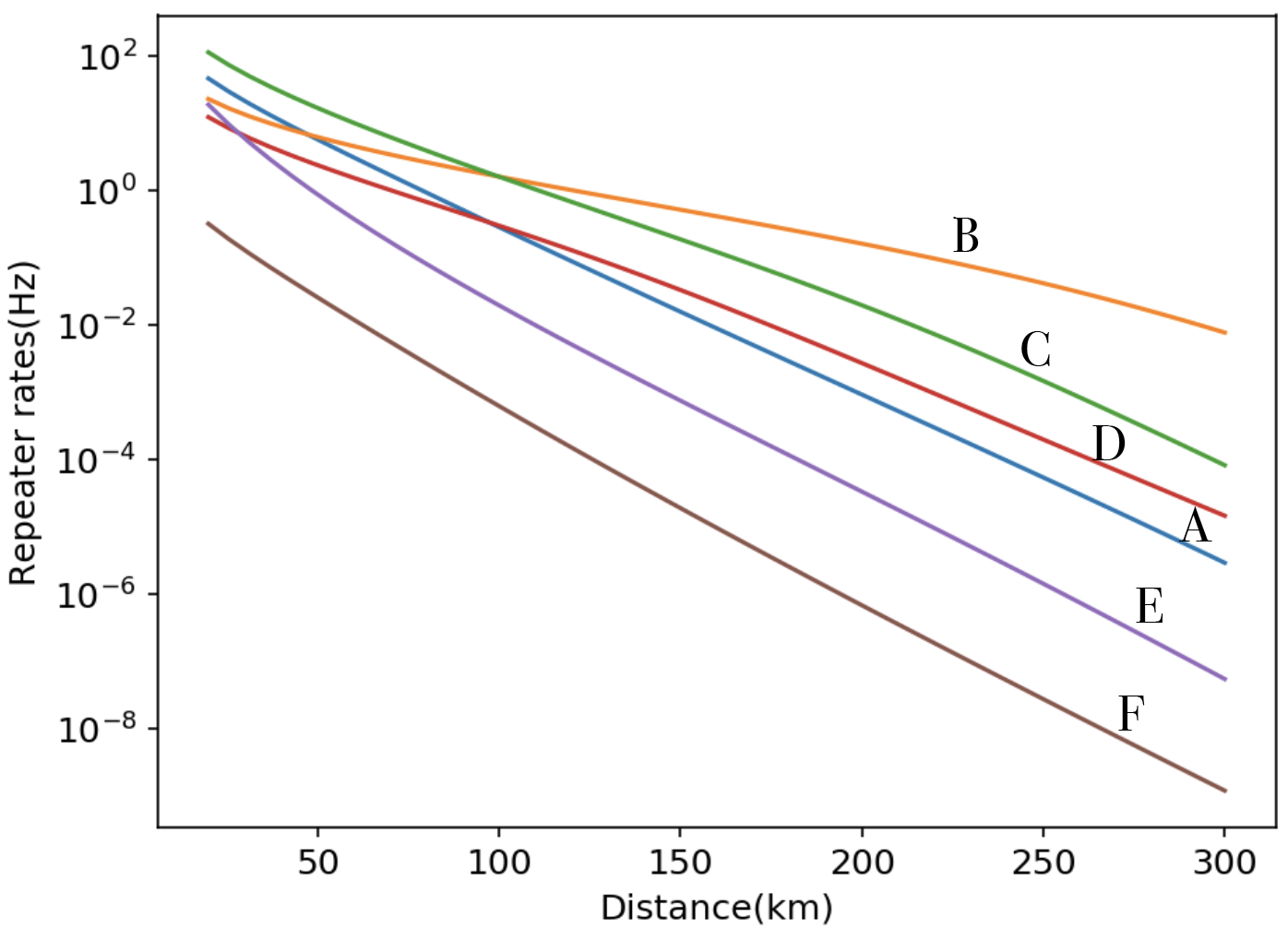}
    \caption{(color online). Comparison of repeater implementations with two links: the ``1 + 1" scheme with QDs and REIs(A); the ``1 + 1" scheme with RAs(B); the ``2 + 2" scheme with QDs and REIs(C); the ``2 + 2" scheme with RAs(D); the ``$\tilde{2}$ + 1" scheme with PDC and REIs(E); the ``$\tilde{2}$ + 2" scheme with PDC and REIs(F)}.
    \label{fig:dis_rate_comp}
\end{figure}
We plot the corresponding two-link repeater performance with various platforms as a function of distance in Fig. \ref{fig:dis_rate_comp}. The solid lines represent the performance of different implementations that are labeled on the figure. The schemes with more deterministic sources achieve higher rates, but even the schemes with non-deterministic sources can allow meaningful proof-of-principle demonstrations, especially the ``$\tilde{2}$ + 1" scheme. Note that the rate for the ``1 + 1'' scheme with RAs decreases more slowly with distance under our assumptions compared with other implementations mainly because of the longer memory lifetime.

\section{Conclusion}
We have studied the near-term performance of different quantum repeater protocols under realistic assumptions, including in particular the effects of exponential memory decay. We reviewed several promising implementations, including the combination of quantum dot sources and rare-earth ion memories as well as Rydberg atom ensembles. Our numerical results can provide useful guidance for the optimization of memory performance in view of long-distance proof-of-principle experiments. Our overall conclusion is that meaningful demonstrations of quantum repeaters with the relatively simple elements considered here are within reach of current technology. Beating direct transmission will likely require further improvements such as multiplexing. 

\section*{Acknowledgement}
We thank Xiaohui Bao for useful discussions and helpful comments on the manuscript. We furthermore acknowledge useful discussions with Jiawei Ji, Faezeh Kimiaee Asadi, and Daniel Oblak. This work was supported by the Natural Sciences and Engineering Research Council (NSERC) of Canada. JianLong Liu was supported by the National Key R\&D Program of China (2017YFA0303902).

\appendix

\section{Detailed derivation of Eq. (\ref{equ:time11}) and Eq. (\ref{equ:time12})}
\label{app:proof}
In this section, we give a detailed derivation of Eq. (\ref{equ:time11}) and Eq. (\ref{equ:time12}). Let us recall the original expression of EDT in Eq. (\ref{equ:time1}), if we substitute $n_{r, max} = n_{r, min} + n_{r, dif}$, we will get
\begin{equation}
    \label{equ:dis_time}
    \begin{split}
       & (\sum_{r = 1}^{k} n_{r, min}T_{0}) p_{k, s}\prod_{r = 1}^{k-1}(1 - p_{r, s}) \\
       + & (\sum_{r = 1}^{k} n_{r, dif}T_{0}) p_{k, s}\prod_{r = 1}^{k-1}(1 - p_{r, s}).
    \end{split}
\end{equation}
To calculate the expectation time, we should average $k$, $n_{r, max}$, and $n_{r, dif}$(contained in $p_{s}$). The calculation for the first term in Eq. (\ref{equ:dis_time}) is easy since $p_{s}$ and $n_{min}$ are independent, while it is hard in the second term where $p_{s}$ depends on $n_{dif}$. Let us first give the expectation value of the first term. As $n_{r, dif}$ is independent with $p_{r, s}$, we derive
\begin{equation}
\begin{split}
\label{equ:app_min}
    & T_{0} \sum_{k = 1}^{\infty} [(\sum_{r = 1}^{k}  \langle n_{r, min} \rangle)
     \langle p_{k, s} \rangle \prod_{r' = 1}^{k-1}(1 - \langle p_{r', s} \rangle) \\
    = & T_{0} \sum_{k = 1}^{\infty}[ k\langle n_{r, min}\rangle  \langle p_{k, s} \rangle \prod_{r' = 1}^{k-1}(1 - \langle p_{r', s} \rangle)]\\
    = &\frac{\langle n_{min}\rangle }{ \langle p_{s} \rangle} T_{0},
\end{split}
\end{equation}
where we have used that the expectation of $n_{min}$ and $p_{s}$ are the same for different swapping attempt, i.e., $\langle n_{r, min}\rangle = \langle n_{min}\rangle$ and $\langle p_{r, s}\rangle = \langle p_{s}\rangle$.

To calculate the expectation value for the second term in Eq. (\ref{equ:dis_time}), we need assumption on the expectation value of $n_{r, dif}p_{r,s}$. As in the main text, we have defined $\beta = \langle n_{dif}p_{s} \rangle / \langle n_{dif}p_{s} \rangle$, where $0 < \beta < 1$. In the upper bound $\beta = 1$, the expectation value of the second term in Eq. (\ref{equ:dis_time}) can be expressed as 

\begin{equation}
\begin{split}
\label{equ:app_dif}
    & T_{0} \sum_{k = 1}^{\infty} [(\sum_{r = 1}^{k}  \langle n_{r, dif} \rangle)
     \langle p_{k, s} \rangle \prod_{r' = 1}^{k-1}(1 - \langle p_{r', s} \rangle) \\
    = & T_{0} \sum_{k = 1}^{\infty}[ k\langle n_{r, dif}\rangle  \langle p_{k, s} \rangle \prod_{r' = 1}^{k-1}(1 - \langle p_{r', s} \rangle)]\\
    = &\frac{\langle n_{dif}\rangle }{ \langle p_{s} \rangle} T_{0},
\end{split}
\end{equation}
while in the lower bound $\beta = 0$, the expectation value of the second term in Eq. (\ref{equ:dis_time}) is negligible compared to the first term. Thus, depends on $\beta$, the average EDT is different. If $\beta = 1$, the average EDT is the sum of Eq. (\ref{equ:app_min}) and Eq. (\ref{equ:app_dif})
\begin{equation}
    \frac{\langle n_{min}\rangle }{ \langle p_{s} \rangle} T_{0} + \frac{\langle n_{dif}\rangle }{ \langle p_{s} \rangle} T_{0} = \frac{\langle n_{max}\rangle }{ \langle p_{s} \rangle} T_{0}.
\end{equation}
On the other hand, if $\beta = 0$, the average EDT is simply the value in Eq. (\ref{equ:app_min})
\begin{equation}
    \frac{\langle n_{min}\rangle }{ \langle p_{s} \rangle} T_{0}.
\end{equation}

\section{Numerical evidence for assumptions}
\label{app:num_evi}
In this section, we give the numerical evidence that supports our assumptions in the main text, especially Eq. (\ref{equ:beta_sps}) and Eq. (\ref{equ:ass_prob_post}). The parameter regime, if not specify, is $p_{0} = 0.01$, $r = T_{0}/\tau_{M} = 1$, $\eta_{d} = 0.95$, and $\alpha^{(0)} = 1$.

\begin{figure*}
    \centering
    \includegraphics[width=\linewidth]{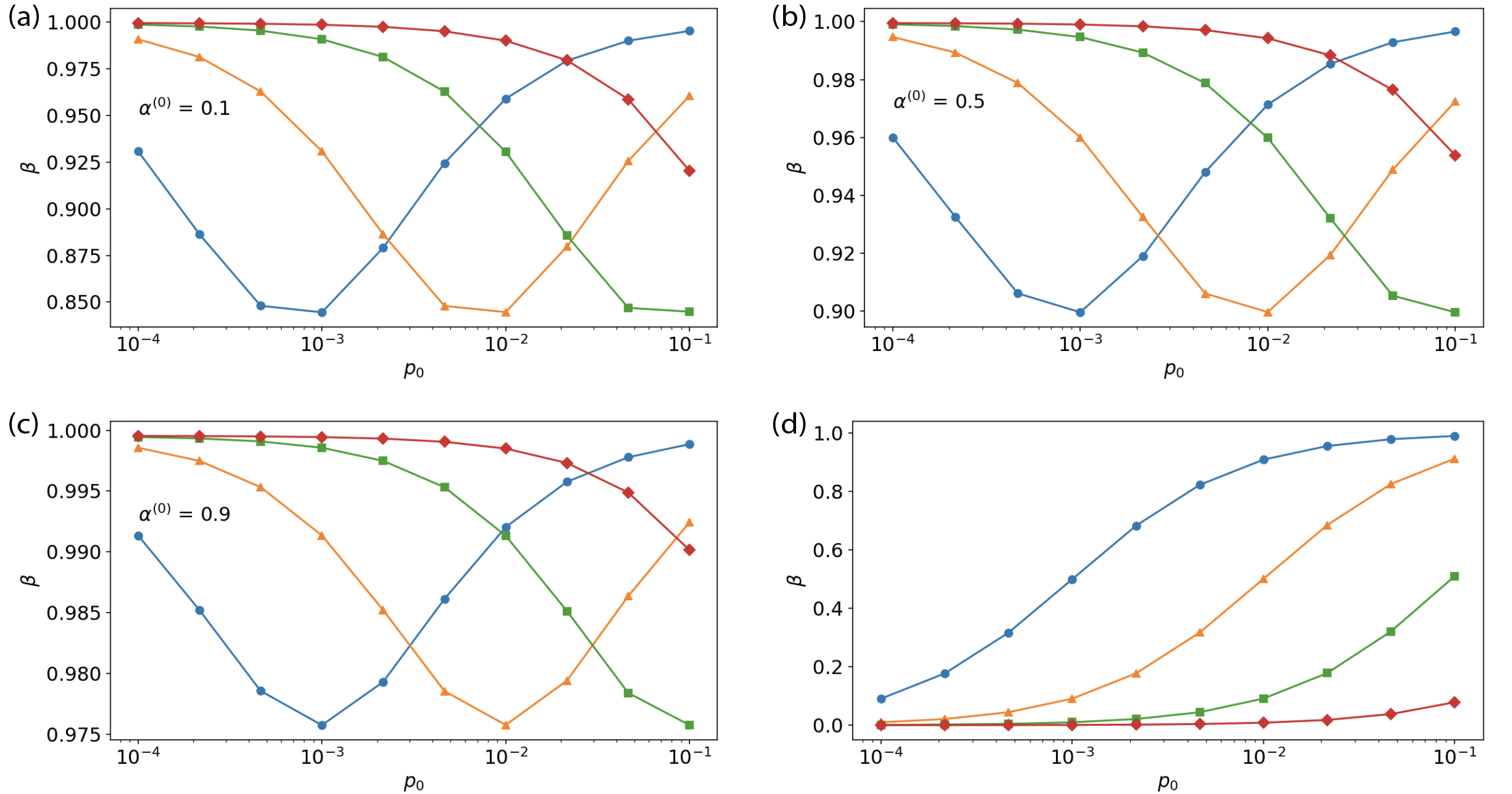}
    \caption{The relation between $\beta$ and $p_{0}$ in different $T_{0}/\tau_{M}$ regime: $T_{0}/\tau_{M} = 0.001$(blue dots), $0.01$(yellow triangles, $0.1$(green squares), and $1$(red diamonds). The first three subfigures shows results for single-photon BSM(a, b and c) and two-photon BSM(d). For single-photon BSM situation, we consider fidelity of entanglement generated state $\alpha^{(0)}$, which depends on the memory efficiency, as 0.9(a), 0.5(b) and 0.1(c). }
    \label{fig:error_rate}
\end{figure*}

First, let us give the numerical evidence for Eq. (\ref{equ:beta_sps}), i.e., $\beta = \langle n_{dif} * p_{s} \rangle / \langle n_{dif}\rangle \langle p_{s} \rangle \approx 1$ in single-photon BSM. We plot $\beta$ as a function of $p_{0}$, shown in Fig.  \ref{fig:error_rate}(a-c) with different choices of state fidelity(after entanglement generation) and different lifetime regime. The result shows that the minimum ratio various with different $p_{0}-t_{M}$ regime and the lowest ratio in our considered regime is $84\%$, which means in Eq. (\ref{equ:beta_sps}), $\beta \approx 1$ is a good approximation. We also plot the ratio in the two-photon BSM scenario: in the low-$p_{0}$ regime, $\beta$ is almost 0, while in the high-$p_{0}$ and high memory lifetime regime, $\beta$ is close to 1. The ratio values 0 and 1 corresponds to the lower bound and upper bound in Eq. (\ref{equ:time12}), respectively, and thus it is easy to approximate the average EDT using the average of lower bound and upper bound. 

\begin{figure*}
    \centering
    \includegraphics[width=\linewidth, height=0.5\linewidth]{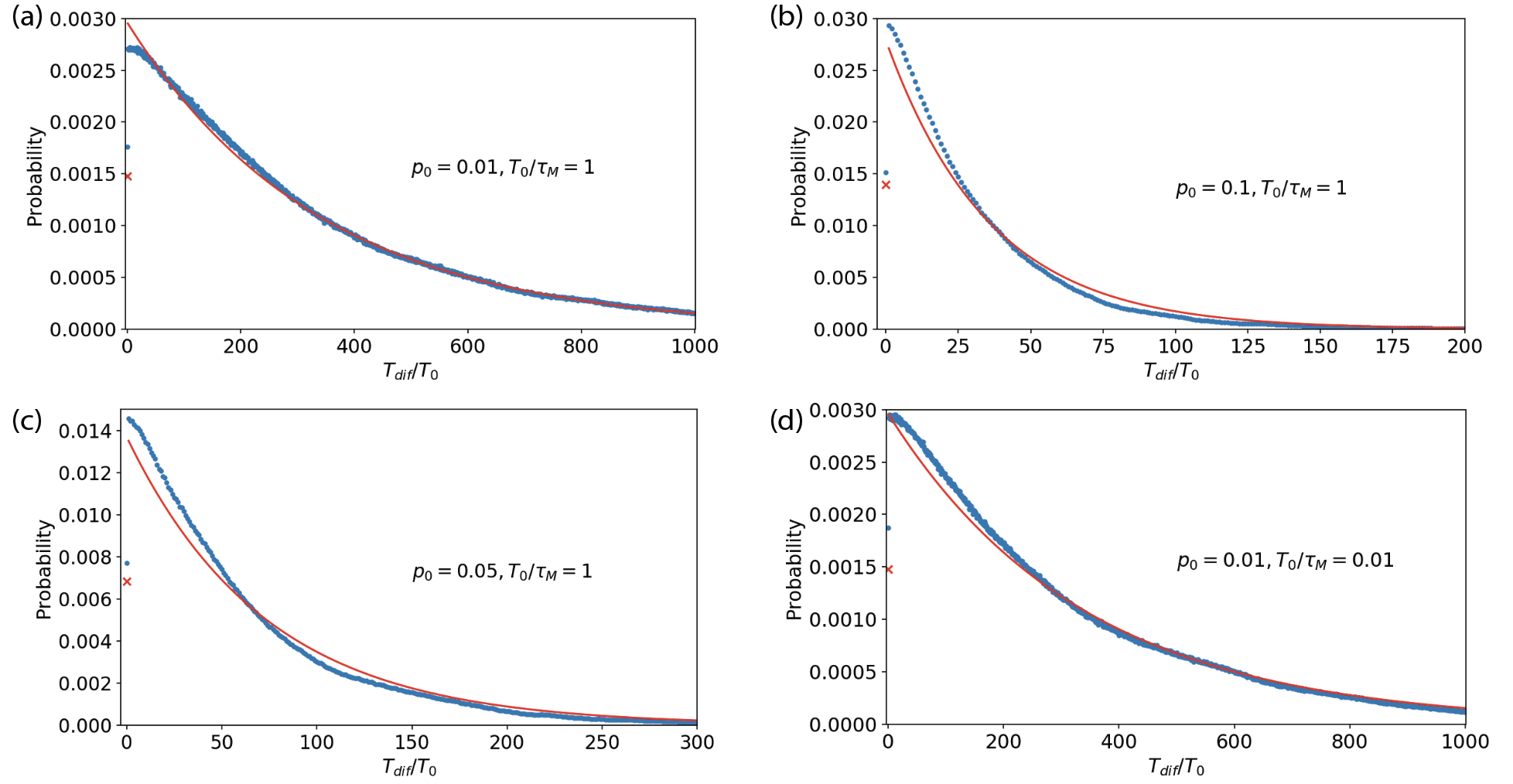}
    \caption{The numerical(blue dot) and theoretical(red solid line) probability distribution for $T_{dif}$. The regime is considered as high-memory-lifetime regime($T_{0}/\tau_{M} = 0.01$), low-memory-lifetime regime($T_{0}/\tau_{M} = 1$), high-$p_{0}$ regime($p_{0} = 0.1$), and low-$p_{0}$ regime($p_{0} = 0.01$). The theoretical prediction fits well with the experimental result.}
    \label{fig:prob_ass}
\end{figure*}

\begin{figure*}[!t]
    \centering
    \includegraphics[width=\linewidth, height=0.5\linewidth]{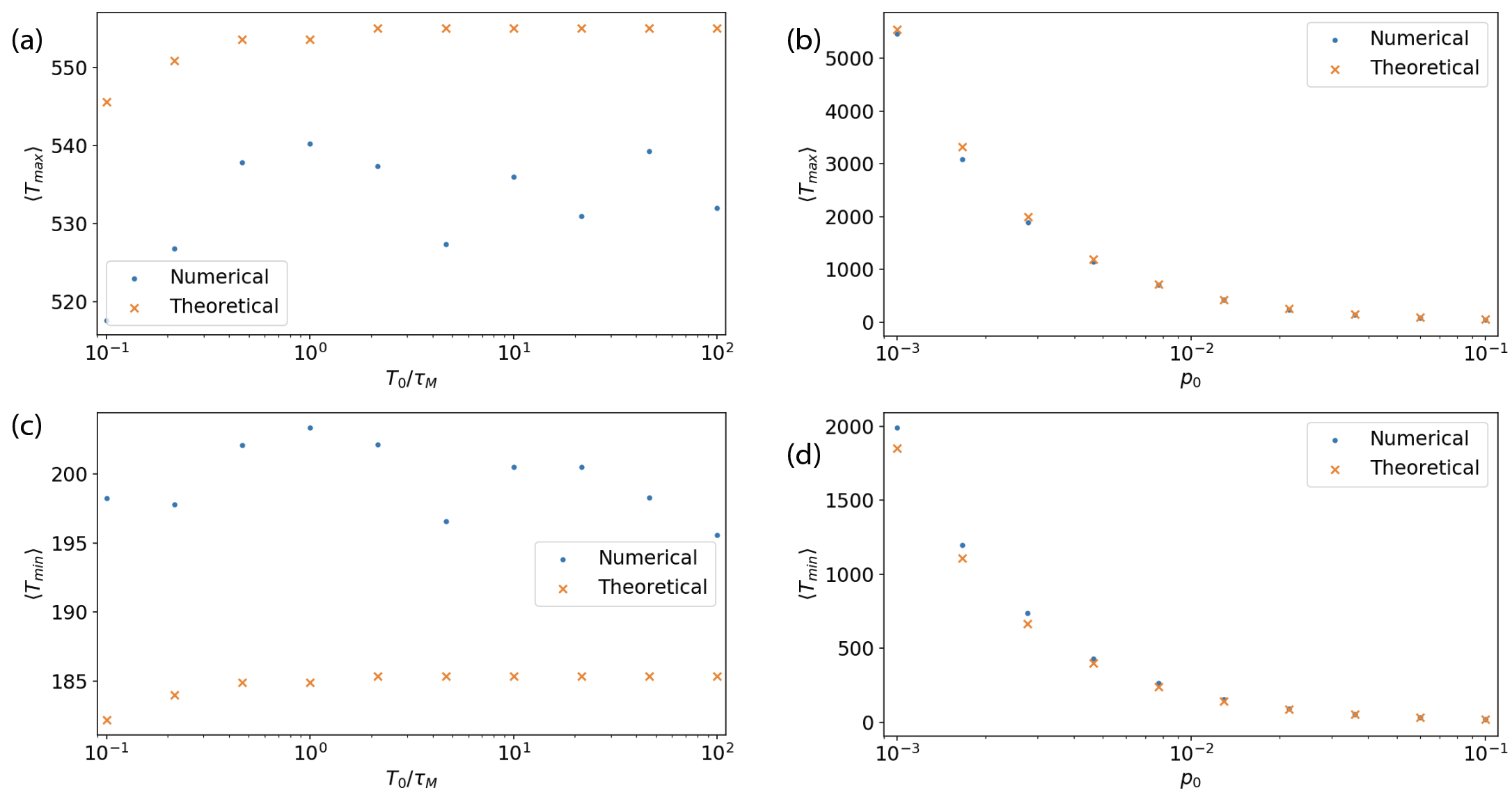}
    \vspace{-1em}
    \caption{The numerical(blue dot) and theoretical(yellow x) expectation value of $T_{min}$(a and b) and $T_{max}$(c and d). We plot the dependence with $T_{0}/\tau_{M}$(a and c) and $p_{0}$(c and d), and find the experimental result and theoretical result fit well.}
    \label{fig:T_max_min}
\end{figure*}

Secondly, we show the assumption in Eq. (\ref{equ:ass_prob_post}) is valid in the approximation of probability distribution of $T_{dif}$ and the expectation value of $T_{min}$ and $T_{max}$. Based on the probability assumption, we plot the theoretical predicted probability distribution of $T_{dif}$ in comparison to the numerical result, shown in Fig. \ref{fig:prob_ass}. The four subfigures correspond to different $p_{0}-t_{M}$ regimes and the theoretical results(red solid line) fit well with the numerical result(blue dots) in all considered parameter regimes. We also compare the theoretically and numerical result of expectation value of $T_{min}$ and $T_{max}$, shown in Fig. \ref{fig:T_max_min}. The theoretical results(blue dots) fit well with the numerical result(yellow cross) for different $\tau_{M}$(subfigures a and c) and $p_{0}$(subfigures b and d).

\section{Comparison of exponential decay and memory cut-off}
\label{app:com}
In this section, we calculate the repeater rates using the memory cut-off assumption, and compare it with our result. Here we focus on the ``2 + 2" scheme.

Given the initial memory state as $\alpha\ket{\Psi}\bra{\Psi} + (1 - \alpha) \rho$, the memory cut-off assumption is that after time $t$, the memory state is  
\begin{equation}
    F = \left\{
             \begin{aligned}
            & \alpha\ket{\Psi}\bra{\Psi} + (1 - \alpha) \rho & t \leq \tau \\
            & \rho' & t > \tau
             \end{aligned}
            \right.,
\end{equation}
where $\rho$ and $\rho'$ are "unwanted state", and $\tau$ is usually the lifetime. It is clear that the fidelity and storage time are negative correlated, and thus the lower bound and upper bound of the average EDT have the same expression as in Eq. (\ref{equ:time12}). We notice the only difference is the average swapping probability $\langle p_{s} \rangle$. Without loss of generality, we assume the prefactor in Eq. (\ref{equ:prob_after2}) as 1, and therefore
\begin{equation}
    p_{s} = \left\{
             \begin{aligned}
            & 1 & t \leq \tau_{M}/2 \\
            & 0 & t > \tau_{M}/2
             \end{aligned}
            \right.,
\end{equation}
where the $\tau_{M}/2$ is because in the ``2 + 2" scheme, both memory will decay. Given $t = n_{dif}T_{0}$, and the probability distribution function is shown in Eq. (\ref{equ:prob_dif}), the expectation value of $p_{s}$ is 
\begin{equation}
    \langle p_{s} \rangle_{cut} = 1 - \frac{2(1 - p_{0})^{\tau_{M}/2T_{0}}}{2 - p_{0}}.
\end{equation}
In comparison, with exponential memory decay, the swapping probability is shown in Eq. (\ref{equ:ps_2+2}), and the expectation value is 
\begin{equation}
    \langle p_{s} \rangle_{exp} = \frac{p_{0}}{2 - p_{0}}\frac{\mathrm{e}^{T_{0}/\tau_{M}} + 1 - p_{0}}{\mathrm{e}^{T_{0}/\tau_{M}} - 1 + p_{0}}.
\end{equation}

\begin{figure}
    \centering
    \includegraphics[width=\linewidth]{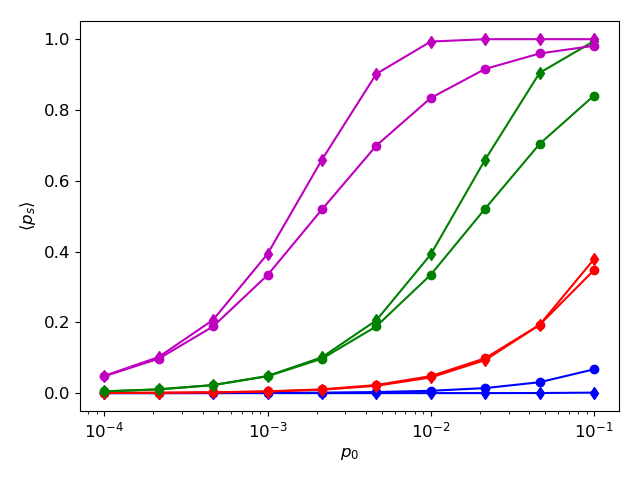}
    \caption{Average swapping probability in memory cut-off assumption $\langle p_{s} \rangle_{cut}$(diamond) and exponential decay assumption $\langle p_{s} \rangle_{exp}$(circle). The plots show the dependence of average swapping probability with different $\tau_{M}/T_{0}$ regime: $\tau_{M}/T_{0} = 1$(blue), $10$(red), $100$(green), and $1000$(magenta).}
    \label{fig:com_exp_cut}
\end{figure}

We plot the average swapping probability under memory cut-off and exponential decay in different $p_{0}$ and $\tau_{M}/T_{0}$ regimes, shown in Fig. \ref{fig:com_exp_cut}. In the high $\tau_{M}/T_{0}$ regimes($>$10), the memory cut-off seems to be a good assumption since the the two expectation values are well matched. However, in the low $\tau_{M}/T_{0}$ regime($\ll$10), we find a distinct difference between the two value. For example, in the regime $\tau_{M}/T_{0} = 1$, which is represented by blue lines in Fig. \ref{fig:com_exp_cut}, $\langle p_{s} \rangle_{cut}/\langle p_{s} \rangle_{exp} = 48.5$ and $521$ corresponding to $p_{0} = 0.1$ and $0.01$. Thus, in low memory time regime($\tau_{M} \sim T_{0}$), the memory cut-off is not a good assumption.

\bibliographystyle{apsrev4-1.bst}
\bibliography{bibgrp}

\begin{thebibliography}{77}%
\makeatletter
\providecommand \@ifxundefined [1]{%
 \@ifx{#1\undefined}
}%
\providecommand \@ifnum [1]{%
 \ifnum #1\expandafter \@firstoftwo
 \else \expandafter \@secondoftwo
 \fi
}%
\providecommand \@ifx [1]{%
 \ifx #1\expandafter \@firstoftwo
 \else \expandafter \@secondoftwo
 \fi
}%
\providecommand \natexlab [1]{#1}%
\providecommand \enquote  [1]{``#1''}%
\providecommand \bibnamefont  [1]{#1}%
\providecommand \bibfnamefont [1]{#1}%
\providecommand \citenamefont [1]{#1}%
\providecommand \href@noop [0]{\@secondoftwo}%
\providecommand \href [0]{\begingroup \@sanitize@url \@href}%
\providecommand \@href[1]{\@@startlink{#1}\@@href}%
\providecommand \@@href[1]{\endgroup#1\@@endlink}%
\providecommand \@sanitize@url [0]{\catcode `\\12\catcode `\$12\catcode
  `\&12\catcode `\#12\catcode `\^12\catcode `\_12\catcode `\%12\relax}%
\providecommand \@@startlink[1]{}%
\providecommand \@@endlink[0]{}%
\providecommand \url  [0]{\begingroup\@sanitize@url \@url }%
\providecommand \@url [1]{\endgroup\@href {#1}{\urlprefix }}%
\providecommand \urlprefix  [0]{URL }%
\providecommand \Eprint [0]{\href }%
\providecommand \doibase [0]{http://dx.doi.org/}%
\providecommand \selectlanguage [0]{\@gobble}%
\providecommand \bibinfo  [0]{\@secondoftwo}%
\providecommand \bibfield  [0]{\@secondoftwo}%
\providecommand \translation [1]{[#1]}%
\providecommand \BibitemOpen [0]{}%
\providecommand \bibitemStop [0]{}%
\providecommand \bibitemNoStop [0]{.\EOS\space}%
\providecommand \EOS [0]{\spacefactor3000\relax}%
\providecommand \BibitemShut  [1]{\csname bibitem#1\endcsname}%
\let\auto@bib@innerbib\@empty
\bibitem [{\citenamefont {Kimble}(2008)}]{kimble2008quantum}%
  \BibitemOpen
  \bibfield  {author} {\bibinfo {author} {\bibfnamefont {H.~J.}\ \bibnamefont
  {Kimble}},\ }\href@noop {} {\bibfield  {journal} {\bibinfo  {journal}
  {Nature}\ }\textbf {\bibinfo {volume} {453}},\ \bibinfo {pages} {1023}
  (\bibinfo {year} {2008})}\BibitemShut {NoStop}%
\bibitem [{\citenamefont {Simon}(2017)}]{simon2017towards}%
  \BibitemOpen
  \bibfield  {author} {\bibinfo {author} {\bibfnamefont {C.}~\bibnamefont
  {Simon}},\ }\href@noop {} {\bibfield  {journal} {\bibinfo  {journal} {Nature
  Photonics}\ }\textbf {\bibinfo {volume} {11}},\ \bibinfo {pages} {678}
  (\bibinfo {year} {2017})}\BibitemShut {NoStop}%
\bibitem [{\citenamefont {Wehner}\ \emph {et~al.}(2018)\citenamefont {Wehner},
  \citenamefont {Elkouss},\ and\ \citenamefont {Hanson}}]{wehner2018quantum}%
  \BibitemOpen
  \bibfield  {author} {\bibinfo {author} {\bibfnamefont {S.}~\bibnamefont
  {Wehner}}, \bibinfo {author} {\bibfnamefont {D.}~\bibnamefont {Elkouss}}, \
  and\ \bibinfo {author} {\bibfnamefont {R.}~\bibnamefont {Hanson}},\
  }\href@noop {} {\bibfield  {journal} {\bibinfo  {journal} {Science}\ }\textbf
  {\bibinfo {volume} {362}},\ \bibinfo {pages} {eaam9288} (\bibinfo {year}
  {2018})}\BibitemShut {NoStop}%
\bibitem [{\citenamefont {Bennett}\ and\ \citenamefont
  {Brassard}(2014)}]{bennett2014quantum}%
  \BibitemOpen
  \bibfield  {author} {\bibinfo {author} {\bibfnamefont {C.~H.}\ \bibnamefont
  {Bennett}}\ and\ \bibinfo {author} {\bibfnamefont {G.}~\bibnamefont
  {Brassard}},\ }\href@noop {} {\bibfield  {journal} {\bibinfo  {journal}
  {Theor. Comput. Sci.}\ }\textbf {\bibinfo {volume} {560}},\ \bibinfo {pages}
  {7} (\bibinfo {year} {2014})}\BibitemShut {NoStop}%
\bibitem [{\citenamefont {Ekert}(1991)}]{ekert1991quantum}%
  \BibitemOpen
  \bibfield  {author} {\bibinfo {author} {\bibfnamefont {A.~K.}\ \bibnamefont
  {Ekert}},\ }\href@noop {} {\bibfield  {journal} {\bibinfo  {journal}
  {Physical review letters}\ }\textbf {\bibinfo {volume} {67}},\ \bibinfo
  {pages} {661} (\bibinfo {year} {1991})}\BibitemShut {NoStop}%
\bibitem [{\citenamefont {Gisin}\ \emph {et~al.}(2002)\citenamefont {Gisin},
  \citenamefont {Ribordy}, \citenamefont {Tittel},\ and\ \citenamefont
  {Zbinden}}]{gisin2002quantum}%
  \BibitemOpen
  \bibfield  {author} {\bibinfo {author} {\bibfnamefont {N.}~\bibnamefont
  {Gisin}}, \bibinfo {author} {\bibfnamefont {G.}~\bibnamefont {Ribordy}},
  \bibinfo {author} {\bibfnamefont {W.}~\bibnamefont {Tittel}}, \ and\ \bibinfo
  {author} {\bibfnamefont {H.}~\bibnamefont {Zbinden}},\ }\href@noop {}
  {\bibfield  {journal} {\bibinfo  {journal} {Reviews of modern physics}\
  }\textbf {\bibinfo {volume} {74}},\ \bibinfo {pages} {145} (\bibinfo {year}
  {2002})}\BibitemShut {NoStop}%
\bibitem [{\citenamefont {Gottesman}\ \emph {et~al.}(2012)\citenamefont
  {Gottesman}, \citenamefont {Jennewein},\ and\ \citenamefont
  {Croke}}]{gottesman2012longer}%
  \BibitemOpen
  \bibfield  {author} {\bibinfo {author} {\bibfnamefont {D.}~\bibnamefont
  {Gottesman}}, \bibinfo {author} {\bibfnamefont {T.}~\bibnamefont
  {Jennewein}}, \ and\ \bibinfo {author} {\bibfnamefont {S.}~\bibnamefont
  {Croke}},\ }\href@noop {} {\bibfield  {journal} {\bibinfo  {journal}
  {Physical review letters}\ }\textbf {\bibinfo {volume} {109}},\ \bibinfo
  {pages} {070503} (\bibinfo {year} {2012})}\BibitemShut {NoStop}%
\bibitem [{\citenamefont {Komar}\ \emph {et~al.}(2014)\citenamefont {Komar},
  \citenamefont {Kessler}, \citenamefont {Bishof}, \citenamefont {Jiang},
  \citenamefont {S{\o}rensen}, \citenamefont {Ye},\ and\ \citenamefont
  {Lukin}}]{komar2014quantum}%
  \BibitemOpen
  \bibfield  {author} {\bibinfo {author} {\bibfnamefont {P.}~\bibnamefont
  {Komar}}, \bibinfo {author} {\bibfnamefont {E.~M.}\ \bibnamefont {Kessler}},
  \bibinfo {author} {\bibfnamefont {M.}~\bibnamefont {Bishof}}, \bibinfo
  {author} {\bibfnamefont {L.}~\bibnamefont {Jiang}}, \bibinfo {author}
  {\bibfnamefont {A.~S.}\ \bibnamefont {S{\o}rensen}}, \bibinfo {author}
  {\bibfnamefont {J.}~\bibnamefont {Ye}}, \ and\ \bibinfo {author}
  {\bibfnamefont {M.~D.}\ \bibnamefont {Lukin}},\ }\href@noop {} {\bibfield
  {journal} {\bibinfo  {journal} {Nature Physics}\ }\textbf {\bibinfo {volume}
  {10}},\ \bibinfo {pages} {582} (\bibinfo {year} {2014})}\BibitemShut
  {NoStop}%
\bibitem [{\citenamefont {Beals}\ \emph {et~al.}(2013)\citenamefont {Beals},
  \citenamefont {Brierley}, \citenamefont {Gray}, \citenamefont {Harrow},
  \citenamefont {Kutin}, \citenamefont {Linden}, \citenamefont {Shepherd},\
  and\ \citenamefont {Stather}}]{beals2013efficient}%
  \BibitemOpen
  \bibfield  {author} {\bibinfo {author} {\bibfnamefont {R.}~\bibnamefont
  {Beals}}, \bibinfo {author} {\bibfnamefont {S.}~\bibnamefont {Brierley}},
  \bibinfo {author} {\bibfnamefont {O.}~\bibnamefont {Gray}}, \bibinfo {author}
  {\bibfnamefont {A.~W.}\ \bibnamefont {Harrow}}, \bibinfo {author}
  {\bibfnamefont {S.}~\bibnamefont {Kutin}}, \bibinfo {author} {\bibfnamefont
  {N.}~\bibnamefont {Linden}}, \bibinfo {author} {\bibfnamefont
  {D.}~\bibnamefont {Shepherd}}, \ and\ \bibinfo {author} {\bibfnamefont
  {M.}~\bibnamefont {Stather}},\ }\href@noop {} {\bibfield  {journal} {\bibinfo
   {journal} {Proceedings of the Royal Society A: Mathematical, Physical and
  Engineering Sciences}\ }\textbf {\bibinfo {volume} {469}},\ \bibinfo {pages}
  {20120686} (\bibinfo {year} {2013})}\BibitemShut {NoStop}%
\bibitem [{\citenamefont {Liao}\ \emph {et~al.}(2017)\citenamefont {Liao},
  \citenamefont {Cai}, \citenamefont {Liu}, \citenamefont {Zhang},
  \citenamefont {Li}, \citenamefont {Ren}, \citenamefont {Yin}, \citenamefont
  {Shen}, \citenamefont {Cao}, \citenamefont {Li} \emph
  {et~al.}}]{liao2017satellite}%
  \BibitemOpen
  \bibfield  {author} {\bibinfo {author} {\bibfnamefont {S.-K.}\ \bibnamefont
  {Liao}}, \bibinfo {author} {\bibfnamefont {W.-Q.}\ \bibnamefont {Cai}},
  \bibinfo {author} {\bibfnamefont {W.-Y.}\ \bibnamefont {Liu}}, \bibinfo
  {author} {\bibfnamefont {L.}~\bibnamefont {Zhang}}, \bibinfo {author}
  {\bibfnamefont {Y.}~\bibnamefont {Li}}, \bibinfo {author} {\bibfnamefont
  {J.-G.}\ \bibnamefont {Ren}}, \bibinfo {author} {\bibfnamefont
  {J.}~\bibnamefont {Yin}}, \bibinfo {author} {\bibfnamefont {Q.}~\bibnamefont
  {Shen}}, \bibinfo {author} {\bibfnamefont {Y.}~\bibnamefont {Cao}}, \bibinfo
  {author} {\bibfnamefont {Z.-P.}\ \bibnamefont {Li}},  \emph {et~al.},\
  }\href@noop {} {\bibfield  {journal} {\bibinfo  {journal} {Nature}\ }\textbf
  {\bibinfo {volume} {549}},\ \bibinfo {pages} {43} (\bibinfo {year}
  {2017})}\BibitemShut {NoStop}%
\bibitem [{\citenamefont {Ren}\ \emph {et~al.}(2017)\citenamefont {Ren},
  \citenamefont {Xu}, \citenamefont {Yong}, \citenamefont {Zhang},
  \citenamefont {Liao}, \citenamefont {Yin}, \citenamefont {Liu}, \citenamefont
  {Cai}, \citenamefont {Yang}, \citenamefont {Li} \emph
  {et~al.}}]{ren2017ground}%
  \BibitemOpen
  \bibfield  {author} {\bibinfo {author} {\bibfnamefont {J.-G.}\ \bibnamefont
  {Ren}}, \bibinfo {author} {\bibfnamefont {P.}~\bibnamefont {Xu}}, \bibinfo
  {author} {\bibfnamefont {H.-L.}\ \bibnamefont {Yong}}, \bibinfo {author}
  {\bibfnamefont {L.}~\bibnamefont {Zhang}}, \bibinfo {author} {\bibfnamefont
  {S.-K.}\ \bibnamefont {Liao}}, \bibinfo {author} {\bibfnamefont
  {J.}~\bibnamefont {Yin}}, \bibinfo {author} {\bibfnamefont {W.-Y.}\
  \bibnamefont {Liu}}, \bibinfo {author} {\bibfnamefont {W.-Q.}\ \bibnamefont
  {Cai}}, \bibinfo {author} {\bibfnamefont {M.}~\bibnamefont {Yang}}, \bibinfo
  {author} {\bibfnamefont {L.}~\bibnamefont {Li}},  \emph {et~al.},\
  }\href@noop {} {\bibfield  {journal} {\bibinfo  {journal} {Nature}\ }\textbf
  {\bibinfo {volume} {549}},\ \bibinfo {pages} {70} (\bibinfo {year}
  {2017})}\BibitemShut {NoStop}%
\bibitem [{\citenamefont {Briegel}\ \emph {et~al.}(1998)\citenamefont
  {Briegel}, \citenamefont {D{\"u}r}, \citenamefont {Cirac},\ and\
  \citenamefont {Zoller}}]{briegel1998quantum}%
  \BibitemOpen
  \bibfield  {author} {\bibinfo {author} {\bibfnamefont {H.-J.}\ \bibnamefont
  {Briegel}}, \bibinfo {author} {\bibfnamefont {W.}~\bibnamefont {D{\"u}r}},
  \bibinfo {author} {\bibfnamefont {J.~I.}\ \bibnamefont {Cirac}}, \ and\
  \bibinfo {author} {\bibfnamefont {P.}~\bibnamefont {Zoller}},\ }\href@noop {}
  {\bibfield  {journal} {\bibinfo  {journal} {Physical Review Letters}\
  }\textbf {\bibinfo {volume} {81}},\ \bibinfo {pages} {5932} (\bibinfo {year}
  {1998})}\BibitemShut {NoStop}%
\bibitem [{\citenamefont {Muralidharan}\ \emph {et~al.}(2016)\citenamefont
  {Muralidharan}, \citenamefont {Li}, \citenamefont {Kim}, \citenamefont
  {L{\"u}tkenhaus}, \citenamefont {Lukin},\ and\ \citenamefont
  {Jiang}}]{muralidharan2016optimal}%
  \BibitemOpen
  \bibfield  {author} {\bibinfo {author} {\bibfnamefont {S.}~\bibnamefont
  {Muralidharan}}, \bibinfo {author} {\bibfnamefont {L.}~\bibnamefont {Li}},
  \bibinfo {author} {\bibfnamefont {J.}~\bibnamefont {Kim}}, \bibinfo {author}
  {\bibfnamefont {N.}~\bibnamefont {L{\"u}tkenhaus}}, \bibinfo {author}
  {\bibfnamefont {M.~D.}\ \bibnamefont {Lukin}}, \ and\ \bibinfo {author}
  {\bibfnamefont {L.}~\bibnamefont {Jiang}},\ }\href@noop {} {\bibfield
  {journal} {\bibinfo  {journal} {Scientific reports}\ }\textbf {\bibinfo
  {volume} {6}},\ \bibinfo {pages} {20463} (\bibinfo {year}
  {2016})}\BibitemShut {NoStop}%
\bibitem [{\citenamefont {Sangouard}\ \emph {et~al.}(2011)\citenamefont
  {Sangouard}, \citenamefont {Simon}, \citenamefont {De~Riedmatten},\ and\
  \citenamefont {Gisin}}]{Sangouard}%
  \BibitemOpen
  \bibfield  {author} {\bibinfo {author} {\bibfnamefont {N.}~\bibnamefont
  {Sangouard}}, \bibinfo {author} {\bibfnamefont {C.}~\bibnamefont {Simon}},
  \bibinfo {author} {\bibfnamefont {H.}~\bibnamefont {De~Riedmatten}}, \ and\
  \bibinfo {author} {\bibfnamefont {N.}~\bibnamefont {Gisin}},\ }\href@noop {}
  {\bibfield  {journal} {\bibinfo  {journal} {Reviews of Modern Physics}\
  }\textbf {\bibinfo {volume} {83}},\ \bibinfo {pages} {33} (\bibinfo {year}
  {2011})}\BibitemShut {NoStop}%
\bibitem [{\citenamefont {Rozp{\k{e}}dek}\ \emph {et~al.}(2019)\citenamefont
  {Rozp{\k{e}}dek}, \citenamefont {Yehia}, \citenamefont {Goodenough},
  \citenamefont {Ruf}, \citenamefont {Humphreys}, \citenamefont {Hanson},
  \citenamefont {Wehner},\ and\ \citenamefont {Elkouss}}]{rozpkedek2019near}%
  \BibitemOpen
  \bibfield  {author} {\bibinfo {author} {\bibfnamefont {F.}~\bibnamefont
  {Rozp{\k{e}}dek}}, \bibinfo {author} {\bibfnamefont {R.}~\bibnamefont
  {Yehia}}, \bibinfo {author} {\bibfnamefont {K.}~\bibnamefont {Goodenough}},
  \bibinfo {author} {\bibfnamefont {M.}~\bibnamefont {Ruf}}, \bibinfo {author}
  {\bibfnamefont {P.~C.}\ \bibnamefont {Humphreys}}, \bibinfo {author}
  {\bibfnamefont {R.}~\bibnamefont {Hanson}}, \bibinfo {author} {\bibfnamefont
  {S.}~\bibnamefont {Wehner}}, \ and\ \bibinfo {author} {\bibfnamefont
  {D.}~\bibnamefont {Elkouss}},\ }\href@noop {} {\bibfield  {journal} {\bibinfo
   {journal} {Physical Review A}\ }\textbf {\bibinfo {volume} {99}},\ \bibinfo
  {pages} {052330} (\bibinfo {year} {2019})}\BibitemShut {NoStop}%
\bibitem [{\citenamefont {Santra}\ \emph {et~al.}(2019)\citenamefont {Santra},
  \citenamefont {Muralidharan}, \citenamefont {Lichtman}, \citenamefont
  {Jiang}, \citenamefont {Monroe},\ and\ \citenamefont
  {Malinovsky}}]{santra2019quantum}%
  \BibitemOpen
  \bibfield  {author} {\bibinfo {author} {\bibfnamefont {S.}~\bibnamefont
  {Santra}}, \bibinfo {author} {\bibfnamefont {S.}~\bibnamefont
  {Muralidharan}}, \bibinfo {author} {\bibfnamefont {M.}~\bibnamefont
  {Lichtman}}, \bibinfo {author} {\bibfnamefont {L.}~\bibnamefont {Jiang}},
  \bibinfo {author} {\bibfnamefont {C.~R.}\ \bibnamefont {Monroe}}, \ and\
  \bibinfo {author} {\bibfnamefont {V.~S.}\ \bibnamefont {Malinovsky}},\
  }\href@noop {} {\bibfield  {journal} {\bibinfo  {journal} {New Journal of
  Physics}\ } (\bibinfo {year} {2019})}\BibitemShut {NoStop}%
\bibitem [{\citenamefont {Krutyanskiy}\ \emph {et~al.}(2019)\citenamefont
  {Krutyanskiy}, \citenamefont {Meraner}, \citenamefont {Schupp}, \citenamefont
  {Krcmarsky}, \citenamefont {Hainzer},\ and\ \citenamefont
  {Lanyon}}]{krutyanskiy2019light}%
  \BibitemOpen
  \bibfield  {author} {\bibinfo {author} {\bibfnamefont {V.}~\bibnamefont
  {Krutyanskiy}}, \bibinfo {author} {\bibfnamefont {M.}~\bibnamefont
  {Meraner}}, \bibinfo {author} {\bibfnamefont {J.}~\bibnamefont {Schupp}},
  \bibinfo {author} {\bibfnamefont {V.}~\bibnamefont {Krcmarsky}}, \bibinfo
  {author} {\bibfnamefont {H.}~\bibnamefont {Hainzer}}, \ and\ \bibinfo
  {author} {\bibfnamefont {B.}~\bibnamefont {Lanyon}},\ }\href@noop {}
  {\bibfield  {journal} {\bibinfo  {journal} {arXiv preprint arXiv:1901.06317}\
  } (\bibinfo {year} {2019})}\BibitemShut {NoStop}%
\bibitem [{\citenamefont {Bao}\ \emph {et~al.}(2012)\citenamefont {Bao},
  \citenamefont {Xu}, \citenamefont {Li}, \citenamefont {Yuan}, \citenamefont
  {Lu},\ and\ \citenamefont {Pan}}]{bao2012quantum}%
  \BibitemOpen
  \bibfield  {author} {\bibinfo {author} {\bibfnamefont {X.-H.}\ \bibnamefont
  {Bao}}, \bibinfo {author} {\bibfnamefont {X.-F.}\ \bibnamefont {Xu}},
  \bibinfo {author} {\bibfnamefont {C.-M.}\ \bibnamefont {Li}}, \bibinfo
  {author} {\bibfnamefont {Z.-S.}\ \bibnamefont {Yuan}}, \bibinfo {author}
  {\bibfnamefont {C.-Y.}\ \bibnamefont {Lu}}, \ and\ \bibinfo {author}
  {\bibfnamefont {J.-W.}\ \bibnamefont {Pan}},\ }\href@noop {} {\bibfield
  {journal} {\bibinfo  {journal} {Proceedings of the National Academy of
  Sciences}\ }\textbf {\bibinfo {volume} {109}},\ \bibinfo {pages} {20347}
  (\bibinfo {year} {2012})}\BibitemShut {NoStop}%
\bibitem [{\citenamefont {Pfaff}\ \emph {et~al.}(2014)\citenamefont {Pfaff},
  \citenamefont {Hensen}, \citenamefont {Bernien}, \citenamefont {van Dam},
  \citenamefont {Blok}, \citenamefont {Taminiau}, \citenamefont {Tiggelman},
  \citenamefont {Schouten}, \citenamefont {Markham}, \citenamefont {Twitchen}
  \emph {et~al.}}]{pfaff2014unconditional}%
  \BibitemOpen
  \bibfield  {author} {\bibinfo {author} {\bibfnamefont {W.}~\bibnamefont
  {Pfaff}}, \bibinfo {author} {\bibfnamefont {B.}~\bibnamefont {Hensen}},
  \bibinfo {author} {\bibfnamefont {H.}~\bibnamefont {Bernien}}, \bibinfo
  {author} {\bibfnamefont {S.~B.}\ \bibnamefont {van Dam}}, \bibinfo {author}
  {\bibfnamefont {M.~S.}\ \bibnamefont {Blok}}, \bibinfo {author}
  {\bibfnamefont {T.~H.}\ \bibnamefont {Taminiau}}, \bibinfo {author}
  {\bibfnamefont {M.~J.}\ \bibnamefont {Tiggelman}}, \bibinfo {author}
  {\bibfnamefont {R.~N.}\ \bibnamefont {Schouten}}, \bibinfo {author}
  {\bibfnamefont {M.}~\bibnamefont {Markham}}, \bibinfo {author} {\bibfnamefont
  {D.~J.}\ \bibnamefont {Twitchen}},  \emph {et~al.},\ }\href@noop {}
  {\bibfield  {journal} {\bibinfo  {journal} {Science}\ }\textbf {\bibinfo
  {volume} {345}},\ \bibinfo {pages} {532} (\bibinfo {year}
  {2014})}\BibitemShut {NoStop}%
\bibitem [{\citenamefont {Vittorini}\ \emph {et~al.}(2014)\citenamefont
  {Vittorini}, \citenamefont {Hucul}, \citenamefont {Inlek}, \citenamefont
  {Crocker},\ and\ \citenamefont {Monroe}}]{vittorini2014entanglement}%
  \BibitemOpen
  \bibfield  {author} {\bibinfo {author} {\bibfnamefont {G.}~\bibnamefont
  {Vittorini}}, \bibinfo {author} {\bibfnamefont {D.}~\bibnamefont {Hucul}},
  \bibinfo {author} {\bibfnamefont {I.}~\bibnamefont {Inlek}}, \bibinfo
  {author} {\bibfnamefont {C.}~\bibnamefont {Crocker}}, \ and\ \bibinfo
  {author} {\bibfnamefont {C.}~\bibnamefont {Monroe}},\ }\href@noop {}
  {\bibfield  {journal} {\bibinfo  {journal} {Physical Review A}\ }\textbf
  {\bibinfo {volume} {90}},\ \bibinfo {pages} {040302} (\bibinfo {year}
  {2014})}\BibitemShut {NoStop}%
\bibitem [{\citenamefont {Humphreys}\ \emph {et~al.}(2018)\citenamefont
  {Humphreys}, \citenamefont {Kalb}, \citenamefont {Morits}, \citenamefont
  {Schouten}, \citenamefont {Vermeulen}, \citenamefont {Twitchen},
  \citenamefont {Markham},\ and\ \citenamefont
  {Hanson}}]{humphreys2018deterministic}%
  \BibitemOpen
  \bibfield  {author} {\bibinfo {author} {\bibfnamefont {P.~C.}\ \bibnamefont
  {Humphreys}}, \bibinfo {author} {\bibfnamefont {N.}~\bibnamefont {Kalb}},
  \bibinfo {author} {\bibfnamefont {J.~P.}\ \bibnamefont {Morits}}, \bibinfo
  {author} {\bibfnamefont {R.~N.}\ \bibnamefont {Schouten}}, \bibinfo {author}
  {\bibfnamefont {R.~F.}\ \bibnamefont {Vermeulen}}, \bibinfo {author}
  {\bibfnamefont {D.~J.}\ \bibnamefont {Twitchen}}, \bibinfo {author}
  {\bibfnamefont {M.}~\bibnamefont {Markham}}, \ and\ \bibinfo {author}
  {\bibfnamefont {R.}~\bibnamefont {Hanson}},\ }\href@noop {} {\bibfield
  {journal} {\bibinfo  {journal} {Nature}\ }\textbf {\bibinfo {volume} {558}},\
  \bibinfo {pages} {268} (\bibinfo {year} {2018})}\BibitemShut {NoStop}%
\bibitem [{\citenamefont {Yu}\ \emph {et~al.}(2019)\citenamefont {Yu},
  \citenamefont {Ma}, \citenamefont {Luo}, \citenamefont {Jing}, \citenamefont
  {Sun}, \citenamefont {Fang}, \citenamefont {Yang}, \citenamefont {Liu},
  \citenamefont {Zheng}, \citenamefont {Xie} \emph
  {et~al.}}]{yu2019entanglement}%
  \BibitemOpen
  \bibfield  {author} {\bibinfo {author} {\bibfnamefont {Y.}~\bibnamefont
  {Yu}}, \bibinfo {author} {\bibfnamefont {F.}~\bibnamefont {Ma}}, \bibinfo
  {author} {\bibfnamefont {X.-Y.}\ \bibnamefont {Luo}}, \bibinfo {author}
  {\bibfnamefont {B.}~\bibnamefont {Jing}}, \bibinfo {author} {\bibfnamefont
  {P.-F.}\ \bibnamefont {Sun}}, \bibinfo {author} {\bibfnamefont {R.-Z.}\
  \bibnamefont {Fang}}, \bibinfo {author} {\bibfnamefont {C.-W.}\ \bibnamefont
  {Yang}}, \bibinfo {author} {\bibfnamefont {H.}~\bibnamefont {Liu}}, \bibinfo
  {author} {\bibfnamefont {M.-Y.}\ \bibnamefont {Zheng}}, \bibinfo {author}
  {\bibfnamefont {X.-P.}\ \bibnamefont {Xie}},  \emph {et~al.},\ }\href@noop {}
  {\bibfield  {journal} {\bibinfo  {journal} {arXiv preprint arXiv:1903.11284}\
  } (\bibinfo {year} {2019})}\BibitemShut {NoStop}%
\bibitem [{\citenamefont {Brask}\ and\ \citenamefont
  {S{\o}rensen}(2008)}]{brask2008memory}%
  \BibitemOpen
  \bibfield  {author} {\bibinfo {author} {\bibfnamefont {J.~B.}\ \bibnamefont
  {Brask}}\ and\ \bibinfo {author} {\bibfnamefont {A.~S.}\ \bibnamefont
  {S{\o}rensen}},\ }\href@noop {} {\bibfield  {journal} {\bibinfo  {journal}
  {Physical Review A}\ }\textbf {\bibinfo {volume} {78}},\ \bibinfo {pages}
  {012350} (\bibinfo {year} {2008})}\BibitemShut {NoStop}%
\bibitem [{\citenamefont {Sinclair}\ \emph {et~al.}(2014)\citenamefont
  {Sinclair}, \citenamefont {Saglamyurek}, \citenamefont {Mallahzadeh},
  \citenamefont {Slater}, \citenamefont {George}, \citenamefont {Ricken},
  \citenamefont {Hedges}, \citenamefont {Oblak}, \citenamefont {Simon},
  \citenamefont {Sohler},\ and\ \citenamefont
  {Tittel}}]{PhysRevLett.113.053603}%
  \BibitemOpen
  \bibfield  {author} {\bibinfo {author} {\bibfnamefont {N.}~\bibnamefont
  {Sinclair}}, \bibinfo {author} {\bibfnamefont {E.}~\bibnamefont
  {Saglamyurek}}, \bibinfo {author} {\bibfnamefont {H.}~\bibnamefont
  {Mallahzadeh}}, \bibinfo {author} {\bibfnamefont {J.~A.}\ \bibnamefont
  {Slater}}, \bibinfo {author} {\bibfnamefont {M.}~\bibnamefont {George}},
  \bibinfo {author} {\bibfnamefont {R.}~\bibnamefont {Ricken}}, \bibinfo
  {author} {\bibfnamefont {M.~P.}\ \bibnamefont {Hedges}}, \bibinfo {author}
  {\bibfnamefont {D.}~\bibnamefont {Oblak}}, \bibinfo {author} {\bibfnamefont
  {C.}~\bibnamefont {Simon}}, \bibinfo {author} {\bibfnamefont
  {W.}~\bibnamefont {Sohler}}, \ and\ \bibinfo {author} {\bibfnamefont
  {W.}~\bibnamefont {Tittel}},\ }\href {\doibase
  10.1103/PhysRevLett.113.053603} {\bibfield  {journal} {\bibinfo  {journal}
  {Phys. Rev. Lett.}\ }\textbf {\bibinfo {volume} {113}},\ \bibinfo {pages}
  {053603} (\bibinfo {year} {2014})}\BibitemShut {NoStop}%
\bibitem [{\citenamefont {Krovi}\ \emph {et~al.}(2016)\citenamefont {Krovi},
  \citenamefont {Guha}, \citenamefont {Dutton}, \citenamefont {Slater},
  \citenamefont {Simon},\ and\ \citenamefont {Tittel}}]{Krovi2016}%
  \BibitemOpen
  \bibfield  {author} {\bibinfo {author} {\bibfnamefont {H.}~\bibnamefont
  {Krovi}}, \bibinfo {author} {\bibfnamefont {S.}~\bibnamefont {Guha}},
  \bibinfo {author} {\bibfnamefont {Z.}~\bibnamefont {Dutton}}, \bibinfo
  {author} {\bibfnamefont {J.~A.}\ \bibnamefont {Slater}}, \bibinfo {author}
  {\bibfnamefont {C.}~\bibnamefont {Simon}}, \ and\ \bibinfo {author}
  {\bibfnamefont {W.}~\bibnamefont {Tittel}},\ }\href {\doibase
  10.1007/s00340-015-6297-4} {\bibfield  {journal} {\bibinfo  {journal}
  {Applied Physics B}\ }\textbf {\bibinfo {volume} {122}},\ \bibinfo {pages}
  {52} (\bibinfo {year} {2016})}\BibitemShut {NoStop}%
\bibitem [{\citenamefont {Collins}\ \emph {et~al.}(2007)\citenamefont
  {Collins}, \citenamefont {Jenkins}, \citenamefont {Kuzmich},\ and\
  \citenamefont {Kennedy}}]{collins2007multiplexed}%
  \BibitemOpen
  \bibfield  {author} {\bibinfo {author} {\bibfnamefont {O.}~\bibnamefont
  {Collins}}, \bibinfo {author} {\bibfnamefont {S.}~\bibnamefont {Jenkins}},
  \bibinfo {author} {\bibfnamefont {A.}~\bibnamefont {Kuzmich}}, \ and\
  \bibinfo {author} {\bibfnamefont {T.}~\bibnamefont {Kennedy}},\ }\href@noop
  {} {\bibfield  {journal} {\bibinfo  {journal} {Physical review letters}\
  }\textbf {\bibinfo {volume} {98}},\ \bibinfo {pages} {060502} (\bibinfo
  {year} {2007})}\BibitemShut {NoStop}%
\bibitem [{\citenamefont {Shchukin}\ \emph {et~al.}(2019)\citenamefont
  {Shchukin}, \citenamefont {Schmidt},\ and\ \citenamefont {van
  Loock}}]{shchukin2019waiting}%
  \BibitemOpen
  \bibfield  {author} {\bibinfo {author} {\bibfnamefont {E.}~\bibnamefont
  {Shchukin}}, \bibinfo {author} {\bibfnamefont {F.}~\bibnamefont {Schmidt}}, \
  and\ \bibinfo {author} {\bibfnamefont {P.}~\bibnamefont {van Loock}},\
  }\href@noop {} {\bibfield  {journal} {\bibinfo  {journal} {Physical Review
  A}\ }\textbf {\bibinfo {volume} {100}},\ \bibinfo {pages} {032322} (\bibinfo
  {year} {2019})}\BibitemShut {NoStop}%
\bibitem [{\citenamefont {Staudt}\ \emph {et~al.}(2007)\citenamefont {Staudt},
  \citenamefont {Afzelius}, \citenamefont {De~Riedmatten}, \citenamefont
  {Hastings-Simon}, \citenamefont {Simon}, \citenamefont {Ricken},
  \citenamefont {Suche}, \citenamefont {Sohler},\ and\ \citenamefont
  {Gisin}}]{staudt2007interference}%
  \BibitemOpen
  \bibfield  {author} {\bibinfo {author} {\bibfnamefont {M.~U.}\ \bibnamefont
  {Staudt}}, \bibinfo {author} {\bibfnamefont {M.}~\bibnamefont {Afzelius}},
  \bibinfo {author} {\bibfnamefont {H.}~\bibnamefont {De~Riedmatten}}, \bibinfo
  {author} {\bibfnamefont {S.~R.}\ \bibnamefont {Hastings-Simon}}, \bibinfo
  {author} {\bibfnamefont {C.}~\bibnamefont {Simon}}, \bibinfo {author}
  {\bibfnamefont {R.}~\bibnamefont {Ricken}}, \bibinfo {author} {\bibfnamefont
  {H.}~\bibnamefont {Suche}}, \bibinfo {author} {\bibfnamefont
  {W.}~\bibnamefont {Sohler}}, \ and\ \bibinfo {author} {\bibfnamefont
  {N.}~\bibnamefont {Gisin}},\ }\href@noop {} {\bibfield  {journal} {\bibinfo
  {journal} {Physical review letters}\ }\textbf {\bibinfo {volume} {99}},\
  \bibinfo {pages} {173602} (\bibinfo {year} {2007})}\BibitemShut {NoStop}%
\bibitem [{\citenamefont {Sangouard}\ \emph {et~al.}(2007)\citenamefont
  {Sangouard}, \citenamefont {Simon}, \citenamefont
  {Min\'a\ifmmode~\check{r}\else \v{r}\fi{}}, \citenamefont {Zbinden},
  \citenamefont {de~Riedmatten},\ and\ \citenamefont
  {Gisin}}]{PhysRevA.76.050301}%
  \BibitemOpen
  \bibfield  {author} {\bibinfo {author} {\bibfnamefont {N.}~\bibnamefont
  {Sangouard}}, \bibinfo {author} {\bibfnamefont {C.}~\bibnamefont {Simon}},
  \bibinfo {author} {\bibfnamefont {J.~c.~v.}\ \bibnamefont
  {Min\'a\ifmmode~\check{r}\else \v{r}\fi{}}}, \bibinfo {author} {\bibfnamefont
  {H.}~\bibnamefont {Zbinden}}, \bibinfo {author} {\bibfnamefont
  {H.}~\bibnamefont {de~Riedmatten}}, \ and\ \bibinfo {author} {\bibfnamefont
  {N.}~\bibnamefont {Gisin}},\ }\href {\doibase 10.1103/PhysRevA.76.050301}
  {\bibfield  {journal} {\bibinfo  {journal} {Phys. Rev. A}\ }\textbf {\bibinfo
  {volume} {76}},\ \bibinfo {pages} {050301} (\bibinfo {year}
  {2007})}\BibitemShut {NoStop}%
\bibitem [{\citenamefont {Simon}\ \emph {et~al.}(2007)\citenamefont {Simon},
  \citenamefont {de~Riedmatten}, \citenamefont {Afzelius}, \citenamefont
  {Sangouard}, \citenamefont {Zbinden},\ and\ \citenamefont
  {Gisin}}]{PhysRevLett.98.190503}%
  \BibitemOpen
  \bibfield  {author} {\bibinfo {author} {\bibfnamefont {C.}~\bibnamefont
  {Simon}}, \bibinfo {author} {\bibfnamefont {H.}~\bibnamefont
  {de~Riedmatten}}, \bibinfo {author} {\bibfnamefont {M.}~\bibnamefont
  {Afzelius}}, \bibinfo {author} {\bibfnamefont {N.}~\bibnamefont {Sangouard}},
  \bibinfo {author} {\bibfnamefont {H.}~\bibnamefont {Zbinden}}, \ and\
  \bibinfo {author} {\bibfnamefont {N.}~\bibnamefont {Gisin}},\ }\href
  {\doibase 10.1103/PhysRevLett.98.190503} {\bibfield  {journal} {\bibinfo
  {journal} {Phys. Rev. Lett.}\ }\textbf {\bibinfo {volume} {98}},\ \bibinfo
  {pages} {190503} (\bibinfo {year} {2007})}\BibitemShut {NoStop}%
\bibitem [{\citenamefont {Siyushev}\ \emph {et~al.}(2014)\citenamefont
  {Siyushev}, \citenamefont {Xia}, \citenamefont {Reuter}, \citenamefont
  {Jamali}, \citenamefont {Zhao}, \citenamefont {Yang}, \citenamefont {Duan},
  \citenamefont {Kukharchyk}, \citenamefont {Wieck}, \citenamefont {Kolesov}
  \emph {et~al.}}]{siyushev2014coherent}%
  \BibitemOpen
  \bibfield  {author} {\bibinfo {author} {\bibfnamefont {P.}~\bibnamefont
  {Siyushev}}, \bibinfo {author} {\bibfnamefont {K.}~\bibnamefont {Xia}},
  \bibinfo {author} {\bibfnamefont {R.}~\bibnamefont {Reuter}}, \bibinfo
  {author} {\bibfnamefont {M.}~\bibnamefont {Jamali}}, \bibinfo {author}
  {\bibfnamefont {N.}~\bibnamefont {Zhao}}, \bibinfo {author} {\bibfnamefont
  {N.}~\bibnamefont {Yang}}, \bibinfo {author} {\bibfnamefont {C.}~\bibnamefont
  {Duan}}, \bibinfo {author} {\bibfnamefont {N.}~\bibnamefont {Kukharchyk}},
  \bibinfo {author} {\bibfnamefont {A.}~\bibnamefont {Wieck}}, \bibinfo
  {author} {\bibfnamefont {R.}~\bibnamefont {Kolesov}},  \emph {et~al.},\
  }\href@noop {} {\bibfield  {journal} {\bibinfo  {journal} {Nature
  communications}\ }\textbf {\bibinfo {volume} {5}},\ \bibinfo {pages} {3895}
  (\bibinfo {year} {2014})}\BibitemShut {NoStop}%
\bibitem [{\citenamefont {Xia}\ \emph {et~al.}(2015)\citenamefont {Xia},
  \citenamefont {Kolesov}, \citenamefont {Wang}, \citenamefont {Siyushev},
  \citenamefont {Reuter}, \citenamefont {Kornher}, \citenamefont {Kukharchyk},
  \citenamefont {Wieck}, \citenamefont {Villa}, \citenamefont {Yang} \emph
  {et~al.}}]{xia2015all}%
  \BibitemOpen
  \bibfield  {author} {\bibinfo {author} {\bibfnamefont {K.}~\bibnamefont
  {Xia}}, \bibinfo {author} {\bibfnamefont {R.}~\bibnamefont {Kolesov}},
  \bibinfo {author} {\bibfnamefont {Y.}~\bibnamefont {Wang}}, \bibinfo {author}
  {\bibfnamefont {P.}~\bibnamefont {Siyushev}}, \bibinfo {author}
  {\bibfnamefont {R.}~\bibnamefont {Reuter}}, \bibinfo {author} {\bibfnamefont
  {T.}~\bibnamefont {Kornher}}, \bibinfo {author} {\bibfnamefont
  {N.}~\bibnamefont {Kukharchyk}}, \bibinfo {author} {\bibfnamefont {A.~D.}\
  \bibnamefont {Wieck}}, \bibinfo {author} {\bibfnamefont {B.}~\bibnamefont
  {Villa}}, \bibinfo {author} {\bibfnamefont {S.}~\bibnamefont {Yang}},  \emph
  {et~al.},\ }\href@noop {} {\bibfield  {journal} {\bibinfo  {journal}
  {Physical review letters}\ }\textbf {\bibinfo {volume} {115}},\ \bibinfo
  {pages} {093602} (\bibinfo {year} {2015})}\BibitemShut {NoStop}%
\bibitem [{\citenamefont {Fraval}\ \emph {et~al.}(2005)\citenamefont {Fraval},
  \citenamefont {Sellars},\ and\ \citenamefont {Longdell}}]{fraval2005dynamic}%
  \BibitemOpen
  \bibfield  {author} {\bibinfo {author} {\bibfnamefont {E.}~\bibnamefont
  {Fraval}}, \bibinfo {author} {\bibfnamefont {M.}~\bibnamefont {Sellars}}, \
  and\ \bibinfo {author} {\bibfnamefont {J.}~\bibnamefont {Longdell}},\
  }\href@noop {} {\bibfield  {journal} {\bibinfo  {journal} {Physical review
  letters}\ }\textbf {\bibinfo {volume} {95}},\ \bibinfo {pages} {030506}
  (\bibinfo {year} {2005})}\BibitemShut {NoStop}%
\bibitem [{\citenamefont {Longdell}\ \emph {et~al.}(2005)\citenamefont
  {Longdell}, \citenamefont {Fraval}, \citenamefont {Sellars},\ and\
  \citenamefont {Manson}}]{longdell2005stopped}%
  \BibitemOpen
  \bibfield  {author} {\bibinfo {author} {\bibfnamefont {J.~J.}\ \bibnamefont
  {Longdell}}, \bibinfo {author} {\bibfnamefont {E.}~\bibnamefont {Fraval}},
  \bibinfo {author} {\bibfnamefont {M.~J.}\ \bibnamefont {Sellars}}, \ and\
  \bibinfo {author} {\bibfnamefont {N.~B.}\ \bibnamefont {Manson}},\
  }\href@noop {} {\bibfield  {journal} {\bibinfo  {journal} {Physical review
  letters}\ }\textbf {\bibinfo {volume} {95}},\ \bibinfo {pages} {063601}
  (\bibinfo {year} {2005})}\BibitemShut {NoStop}%
\bibitem [{\citenamefont {Zhong}\ \emph {et~al.}(2015)\citenamefont {Zhong},
  \citenamefont {Hedges}, \citenamefont {Ahlefeldt}, \citenamefont
  {Bartholomew}, \citenamefont {Beavan}, \citenamefont {Wittig}, \citenamefont
  {Longdell},\ and\ \citenamefont {Sellars}}]{Zhong2015}%
  \BibitemOpen
  \bibfield  {author} {\bibinfo {author} {\bibfnamefont {M.}~\bibnamefont
  {Zhong}}, \bibinfo {author} {\bibfnamefont {M.~P.}\ \bibnamefont {Hedges}},
  \bibinfo {author} {\bibfnamefont {R.~L.}\ \bibnamefont {Ahlefeldt}}, \bibinfo
  {author} {\bibfnamefont {J.~G.}\ \bibnamefont {Bartholomew}}, \bibinfo
  {author} {\bibfnamefont {S.~E.}\ \bibnamefont {Beavan}}, \bibinfo {author}
  {\bibfnamefont {S.~M.}\ \bibnamefont {Wittig}}, \bibinfo {author}
  {\bibfnamefont {J.~J.}\ \bibnamefont {Longdell}}, \ and\ \bibinfo {author}
  {\bibfnamefont {M.~J.}\ \bibnamefont {Sellars}},\ }\href {\doibase
  10.1038/nature14025} {\bibfield  {journal} {\bibinfo  {journal} {Nature}\
  }\textbf {\bibinfo {volume} {517}},\ \bibinfo {pages} {177} (\bibinfo {year}
  {2015})}\BibitemShut {NoStop}%
\bibitem [{\citenamefont {Ran{\v{c}}i{\'{c}}}\ \emph
  {et~al.}(2017)\citenamefont {Ran{\v{c}}i{\'{c}}}, \citenamefont {Hedges},
  \citenamefont {Ahlefeldt},\ and\ \citenamefont {Sellars}}]{Rancic2017}%
  \BibitemOpen
  \bibfield  {author} {\bibinfo {author} {\bibfnamefont {M.}~\bibnamefont
  {Ran{\v{c}}i{\'{c}}}}, \bibinfo {author} {\bibfnamefont {M.~P.}\ \bibnamefont
  {Hedges}}, \bibinfo {author} {\bibfnamefont {R.~L.}\ \bibnamefont
  {Ahlefeldt}}, \ and\ \bibinfo {author} {\bibfnamefont {M.~J.}\ \bibnamefont
  {Sellars}},\ }\href {\doibase 10.1038/nphys4254} {\bibfield  {journal}
  {\bibinfo  {journal} {Nature Physics}\ }\textbf {\bibinfo {volume} {14}},\
  \bibinfo {pages} {50} (\bibinfo {year} {2017})}\BibitemShut {NoStop}%
\bibitem [{\citenamefont {Sabooni}\ \emph {et~al.}(2010)\citenamefont
  {Sabooni}, \citenamefont {Beaudoin}, \citenamefont {Walther}, \citenamefont
  {Lin}, \citenamefont {Amari}, \citenamefont {Huang},\ and\ \citenamefont
  {Kr{\"o}ll}}]{sabooni2010storage}%
  \BibitemOpen
  \bibfield  {author} {\bibinfo {author} {\bibfnamefont {M.}~\bibnamefont
  {Sabooni}}, \bibinfo {author} {\bibfnamefont {F.}~\bibnamefont {Beaudoin}},
  \bibinfo {author} {\bibfnamefont {A.}~\bibnamefont {Walther}}, \bibinfo
  {author} {\bibfnamefont {N.}~\bibnamefont {Lin}}, \bibinfo {author}
  {\bibfnamefont {A.}~\bibnamefont {Amari}}, \bibinfo {author} {\bibfnamefont
  {M.}~\bibnamefont {Huang}}, \ and\ \bibinfo {author} {\bibfnamefont
  {S.}~\bibnamefont {Kr{\"o}ll}},\ }\href@noop {} {\bibfield  {journal}
  {\bibinfo  {journal} {Physical review letters}\ }\textbf {\bibinfo {volume}
  {105}},\ \bibinfo {pages} {060501} (\bibinfo {year} {2010})}\BibitemShut
  {NoStop}%
\bibitem [{\citenamefont {Usmani}\ \emph {et~al.}(2010)\citenamefont {Usmani},
  \citenamefont {Afzelius}, \citenamefont {De~Riedmatten},\ and\ \citenamefont
  {Gisin}}]{usmani2010mapping}%
  \BibitemOpen
  \bibfield  {author} {\bibinfo {author} {\bibfnamefont {I.}~\bibnamefont
  {Usmani}}, \bibinfo {author} {\bibfnamefont {M.}~\bibnamefont {Afzelius}},
  \bibinfo {author} {\bibfnamefont {H.}~\bibnamefont {De~Riedmatten}}, \ and\
  \bibinfo {author} {\bibfnamefont {N.}~\bibnamefont {Gisin}},\ }\href@noop {}
  {\bibfield  {journal} {\bibinfo  {journal} {Nature Communications}\ }\textbf
  {\bibinfo {volume} {1}},\ \bibinfo {pages} {12} (\bibinfo {year}
  {2010})}\BibitemShut {NoStop}%
\bibitem [{\citenamefont {Amari}\ \emph {et~al.}(2010)\citenamefont {Amari},
  \citenamefont {Walther}, \citenamefont {Sabooni}, \citenamefont {Huang},
  \citenamefont {Kr{\"{o}}ll}, \citenamefont {Afzelius}, \citenamefont
  {Usmani}, \citenamefont {Lauritzen}, \citenamefont {Sangouard}, \citenamefont
  {de~Riedmatten},\ and\ \citenamefont {Gisin}}]{AMARI20101579}%
  \BibitemOpen
  \bibfield  {author} {\bibinfo {author} {\bibfnamefont {A.}~\bibnamefont
  {Amari}}, \bibinfo {author} {\bibfnamefont {A.}~\bibnamefont {Walther}},
  \bibinfo {author} {\bibfnamefont {M.}~\bibnamefont {Sabooni}}, \bibinfo
  {author} {\bibfnamefont {M.}~\bibnamefont {Huang}}, \bibinfo {author}
  {\bibfnamefont {S.}~\bibnamefont {Kr{\"{o}}ll}}, \bibinfo {author}
  {\bibfnamefont {M.}~\bibnamefont {Afzelius}}, \bibinfo {author}
  {\bibfnamefont {I.}~\bibnamefont {Usmani}}, \bibinfo {author} {\bibfnamefont
  {B.}~\bibnamefont {Lauritzen}}, \bibinfo {author} {\bibfnamefont
  {N.}~\bibnamefont {Sangouard}}, \bibinfo {author} {\bibfnamefont
  {H.}~\bibnamefont {de~Riedmatten}}, \ and\ \bibinfo {author} {\bibfnamefont
  {N.}~\bibnamefont {Gisin}},\ }\href {\doibase
  https://doi.org/10.1016/j.jlumin.2010.01.012} {\bibfield  {journal} {\bibinfo
   {journal} {Journal of Luminescence}\ }\textbf {\bibinfo {volume} {130}},\
  \bibinfo {pages} {1579} (\bibinfo {year} {2010})}\BibitemShut {NoStop}%
\bibitem [{\citenamefont {Bonarota}\ \emph {et~al.}(2011)\citenamefont
  {Bonarota}, \citenamefont {Gou{\"{e}}t},\ and\ \citenamefont
  {Chaneli{\`{e}}re}}]{Bonarota_2011}%
  \BibitemOpen
  \bibfield  {author} {\bibinfo {author} {\bibfnamefont {M.}~\bibnamefont
  {Bonarota}}, \bibinfo {author} {\bibfnamefont {J.-L.~L.}\ \bibnamefont
  {Gou{\"{e}}t}}, \ and\ \bibinfo {author} {\bibfnamefont {T.}~\bibnamefont
  {Chaneli{\`{e}}re}},\ }\href {\doibase 10.1088/1367-2630/13/1/013013}
  {\bibfield  {journal} {\bibinfo  {journal} {New Journal of Physics}\ }\textbf
  {\bibinfo {volume} {13}},\ \bibinfo {pages} {13013} (\bibinfo {year}
  {2011})}\BibitemShut {NoStop}%
\bibitem [{\citenamefont {Timoney}\ \emph {et~al.}(2012)\citenamefont
  {Timoney}, \citenamefont {Lauritzen}, \citenamefont {Usmani}, \citenamefont
  {Afzelius},\ and\ \citenamefont {Gisin}}]{timoney2012atomic}%
  \BibitemOpen
  \bibfield  {author} {\bibinfo {author} {\bibfnamefont {N.}~\bibnamefont
  {Timoney}}, \bibinfo {author} {\bibfnamefont {B.}~\bibnamefont {Lauritzen}},
  \bibinfo {author} {\bibfnamefont {I.}~\bibnamefont {Usmani}}, \bibinfo
  {author} {\bibfnamefont {M.}~\bibnamefont {Afzelius}}, \ and\ \bibinfo
  {author} {\bibfnamefont {N.}~\bibnamefont {Gisin}},\ }\href@noop {}
  {\bibfield  {journal} {\bibinfo  {journal} {Journal of Physics B: Atomic,
  Molecular and Optical Physics}\ }\textbf {\bibinfo {volume} {45}},\ \bibinfo
  {pages} {124001} (\bibinfo {year} {2012})}\BibitemShut {NoStop}%
\bibitem [{\citenamefont {Jobez}\ \emph {et~al.}(2014)\citenamefont {Jobez},
  \citenamefont {Usmani}, \citenamefont {Timoney}, \citenamefont {Laplane},
  \citenamefont {Gisin},\ and\ \citenamefont {Afzelius}}]{Jobez_2014}%
  \BibitemOpen
  \bibfield  {author} {\bibinfo {author} {\bibfnamefont {P.}~\bibnamefont
  {Jobez}}, \bibinfo {author} {\bibfnamefont {I.}~\bibnamefont {Usmani}},
  \bibinfo {author} {\bibfnamefont {N.}~\bibnamefont {Timoney}}, \bibinfo
  {author} {\bibfnamefont {C.}~\bibnamefont {Laplane}}, \bibinfo {author}
  {\bibfnamefont {N.}~\bibnamefont {Gisin}}, \ and\ \bibinfo {author}
  {\bibfnamefont {M.}~\bibnamefont {Afzelius}},\ }\href {\doibase
  10.1088/1367-2630/16/8/083005} {\bibfield  {journal} {\bibinfo  {journal}
  {New Journal of Physics}\ }\textbf {\bibinfo {volume} {16}},\ \bibinfo
  {pages} {83005} (\bibinfo {year} {2014})}\BibitemShut {NoStop}%
\bibitem [{\citenamefont {Sabooni}\ \emph {et~al.}(2013)\citenamefont
  {Sabooni}, \citenamefont {Li}, \citenamefont {Kr{\"o}ll},\ and\ \citenamefont
  {Rippe}}]{sabooni2013efficient}%
  \BibitemOpen
  \bibfield  {author} {\bibinfo {author} {\bibfnamefont {M.}~\bibnamefont
  {Sabooni}}, \bibinfo {author} {\bibfnamefont {Q.}~\bibnamefont {Li}},
  \bibinfo {author} {\bibfnamefont {S.}~\bibnamefont {Kr{\"o}ll}}, \ and\
  \bibinfo {author} {\bibfnamefont {L.}~\bibnamefont {Rippe}},\ }\href@noop {}
  {\bibfield  {journal} {\bibinfo  {journal} {Physical review letters}\
  }\textbf {\bibinfo {volume} {110}},\ \bibinfo {pages} {133604} (\bibinfo
  {year} {2013})}\BibitemShut {NoStop}%
\bibitem [{\citenamefont {Lukin}\ \emph {et~al.}(2001)\citenamefont {Lukin},
  \citenamefont {Fleischhauer}, \citenamefont {Cote}, \citenamefont {Duan},
  \citenamefont {Jaksch}, \citenamefont {Cirac},\ and\ \citenamefont
  {Zoller}}]{PhysRevLett.87.037901}%
  \BibitemOpen
  \bibfield  {author} {\bibinfo {author} {\bibfnamefont {M.~D.}\ \bibnamefont
  {Lukin}}, \bibinfo {author} {\bibfnamefont {M.}~\bibnamefont {Fleischhauer}},
  \bibinfo {author} {\bibfnamefont {R.}~\bibnamefont {Cote}}, \bibinfo {author}
  {\bibfnamefont {L.~M.}\ \bibnamefont {Duan}}, \bibinfo {author}
  {\bibfnamefont {D.}~\bibnamefont {Jaksch}}, \bibinfo {author} {\bibfnamefont
  {J.~I.}\ \bibnamefont {Cirac}}, \ and\ \bibinfo {author} {\bibfnamefont
  {P.}~\bibnamefont {Zoller}},\ }\href {\doibase 10.1103/PhysRevLett.87.037901}
  {\bibfield  {journal} {\bibinfo  {journal} {Phys. Rev. Lett.}\ }\textbf
  {\bibinfo {volume} {87}},\ \bibinfo {pages} {037901} (\bibinfo {year}
  {2001})}\BibitemShut {NoStop}%
\bibitem [{\citenamefont {Saffman}\ and\ \citenamefont
  {Walker}(2002)}]{saffman2002creating}%
  \BibitemOpen
  \bibfield  {author} {\bibinfo {author} {\bibfnamefont {M.}~\bibnamefont
  {Saffman}}\ and\ \bibinfo {author} {\bibfnamefont {T.}~\bibnamefont
  {Walker}},\ }\href@noop {} {\bibfield  {journal} {\bibinfo  {journal}
  {Physical Review A}\ }\textbf {\bibinfo {volume} {66}},\ \bibinfo {pages}
  {065403} (\bibinfo {year} {2002})}\BibitemShut {NoStop}%
\bibitem [{\citenamefont {de~L\'es\'eleuc}\ \emph {et~al.}(2018)\citenamefont
  {de~L\'es\'eleuc}, \citenamefont {Barredo}, \citenamefont {Lienhard},
  \citenamefont {Browaeys},\ and\ \citenamefont {Lahaye}}]{PhysRevA.97.053803}%
  \BibitemOpen
  \bibfield  {author} {\bibinfo {author} {\bibfnamefont {S.}~\bibnamefont
  {de~L\'es\'eleuc}}, \bibinfo {author} {\bibfnamefont {D.}~\bibnamefont
  {Barredo}}, \bibinfo {author} {\bibfnamefont {V.}~\bibnamefont {Lienhard}},
  \bibinfo {author} {\bibfnamefont {A.}~\bibnamefont {Browaeys}}, \ and\
  \bibinfo {author} {\bibfnamefont {T.}~\bibnamefont {Lahaye}},\ }\href
  {\doibase 10.1103/PhysRevA.97.053803} {\bibfield  {journal} {\bibinfo
  {journal} {Phys. Rev. A}\ }\textbf {\bibinfo {volume} {97}},\ \bibinfo
  {pages} {053803} (\bibinfo {year} {2018})}\BibitemShut {NoStop}%
\bibitem [{\citenamefont {Li}\ \emph {et~al.}(2019{\natexlab{a}})\citenamefont
  {Li}, \citenamefont {Zhou}, \citenamefont {Yang}, \citenamefont {Sun},
  \citenamefont {Liu}, \citenamefont {Bao},\ and\ \citenamefont
  {Pan}}]{PhysRevLett.123.140504}%
  \BibitemOpen
  \bibfield  {author} {\bibinfo {author} {\bibfnamefont {J.}~\bibnamefont
  {Li}}, \bibinfo {author} {\bibfnamefont {M.-T.}\ \bibnamefont {Zhou}},
  \bibinfo {author} {\bibfnamefont {C.-W.}\ \bibnamefont {Yang}}, \bibinfo
  {author} {\bibfnamefont {P.-F.}\ \bibnamefont {Sun}}, \bibinfo {author}
  {\bibfnamefont {J.-L.}\ \bibnamefont {Liu}}, \bibinfo {author} {\bibfnamefont
  {X.-H.}\ \bibnamefont {Bao}}, \ and\ \bibinfo {author} {\bibfnamefont
  {J.-W.}\ \bibnamefont {Pan}},\ }\href {\doibase
  10.1103/PhysRevLett.123.140504} {\bibfield  {journal} {\bibinfo  {journal}
  {Phys. Rev. Lett.}\ }\textbf {\bibinfo {volume} {123}},\ \bibinfo {pages}
  {140504} (\bibinfo {year} {2019}{\natexlab{a}})}\BibitemShut {NoStop}%
\bibitem [{\citenamefont {Yang}\ \emph {et~al.}(2016)\citenamefont {Yang},
  \citenamefont {Wang}, \citenamefont {Bao},\ and\ \citenamefont
  {Pan}}]{yang2016efficient}%
  \BibitemOpen
  \bibfield  {author} {\bibinfo {author} {\bibfnamefont {S.-J.}\ \bibnamefont
  {Yang}}, \bibinfo {author} {\bibfnamefont {X.-J.}\ \bibnamefont {Wang}},
  \bibinfo {author} {\bibfnamefont {X.-H.}\ \bibnamefont {Bao}}, \ and\
  \bibinfo {author} {\bibfnamefont {J.-W.}\ \bibnamefont {Pan}},\ }\href@noop
  {} {\bibfield  {journal} {\bibinfo  {journal} {Nature Photonics}\ }\textbf
  {\bibinfo {volume} {10}},\ \bibinfo {pages} {381} (\bibinfo {year}
  {2016})}\BibitemShut {NoStop}%
\bibitem [{\citenamefont {Couteau}(2018)}]{Couteau_2018}%
  \BibitemOpen
  \bibfield  {author} {\bibinfo {author} {\bibfnamefont {C.}~\bibnamefont
  {Couteau}},\ }\href {\doibase 10.1080/00107514.2018.1488463} {\bibfield
  {journal} {\bibinfo  {journal} {Contemporary Physics}\ }\textbf {\bibinfo
  {volume} {59}},\ \bibinfo {pages} {291–304} (\bibinfo {year}
  {2018})}\BibitemShut {NoStop}%
\bibitem [{\citenamefont {Senellart}\ \emph {et~al.}(2017)\citenamefont
  {Senellart}, \citenamefont {Solomon},\ and\ \citenamefont
  {White}}]{senellart2017high}%
  \BibitemOpen
  \bibfield  {author} {\bibinfo {author} {\bibfnamefont {P.}~\bibnamefont
  {Senellart}}, \bibinfo {author} {\bibfnamefont {G.}~\bibnamefont {Solomon}},
  \ and\ \bibinfo {author} {\bibfnamefont {A.}~\bibnamefont {White}},\
  }\href@noop {} {\bibfield  {journal} {\bibinfo  {journal} {Nature
  nanotechnology}\ }\textbf {\bibinfo {volume} {12}},\ \bibinfo {pages} {1026}
  (\bibinfo {year} {2017})}\BibitemShut {NoStop}%
\bibitem [{\citenamefont {Ding}\ \emph {et~al.}(2016)\citenamefont {Ding},
  \citenamefont {He}, \citenamefont {Duan}, \citenamefont {Gregersen},
  \citenamefont {Chen}, \citenamefont {Unsleber}, \citenamefont {Maier},
  \citenamefont {Schneider}, \citenamefont {Kamp}, \citenamefont {H{\"o}fling}
  \emph {et~al.}}]{ding2016demand}%
  \BibitemOpen
  \bibfield  {author} {\bibinfo {author} {\bibfnamefont {X.}~\bibnamefont
  {Ding}}, \bibinfo {author} {\bibfnamefont {Y.}~\bibnamefont {He}}, \bibinfo
  {author} {\bibfnamefont {Z.-C.}\ \bibnamefont {Duan}}, \bibinfo {author}
  {\bibfnamefont {N.}~\bibnamefont {Gregersen}}, \bibinfo {author}
  {\bibfnamefont {M.-C.}\ \bibnamefont {Chen}}, \bibinfo {author}
  {\bibfnamefont {S.}~\bibnamefont {Unsleber}}, \bibinfo {author}
  {\bibfnamefont {S.}~\bibnamefont {Maier}}, \bibinfo {author} {\bibfnamefont
  {C.}~\bibnamefont {Schneider}}, \bibinfo {author} {\bibfnamefont
  {M.}~\bibnamefont {Kamp}}, \bibinfo {author} {\bibfnamefont {S.}~\bibnamefont
  {H{\"o}fling}},  \emph {et~al.},\ }\href@noop {} {\bibfield  {journal}
  {\bibinfo  {journal} {Physical review letters}\ }\textbf {\bibinfo {volume}
  {116}},\ \bibinfo {pages} {020401} (\bibinfo {year} {2016})}\BibitemShut
  {NoStop}%
\bibitem [{\citenamefont {Wang}\ \emph {et~al.}(2019)\citenamefont {Wang},
  \citenamefont {Hu}, \citenamefont {Chung}, \citenamefont {Qin}, \citenamefont
  {Yang}, \citenamefont {Li}, \citenamefont {Liu}, \citenamefont {Zhong},
  \citenamefont {He}, \citenamefont {Ding} \emph {et~al.}}]{wang2019demand}%
  \BibitemOpen
  \bibfield  {author} {\bibinfo {author} {\bibfnamefont {H.}~\bibnamefont
  {Wang}}, \bibinfo {author} {\bibfnamefont {H.}~\bibnamefont {Hu}}, \bibinfo
  {author} {\bibfnamefont {T.-H.}\ \bibnamefont {Chung}}, \bibinfo {author}
  {\bibfnamefont {J.}~\bibnamefont {Qin}}, \bibinfo {author} {\bibfnamefont
  {X.}~\bibnamefont {Yang}}, \bibinfo {author} {\bibfnamefont {J.-P.}\
  \bibnamefont {Li}}, \bibinfo {author} {\bibfnamefont {R.-Z.}\ \bibnamefont
  {Liu}}, \bibinfo {author} {\bibfnamefont {H.-S.}\ \bibnamefont {Zhong}},
  \bibinfo {author} {\bibfnamefont {Y.-M.}\ \bibnamefont {He}}, \bibinfo
  {author} {\bibfnamefont {X.}~\bibnamefont {Ding}},  \emph {et~al.},\
  }\href@noop {} {\bibfield  {journal} {\bibinfo  {journal} {Physical Review
  Letters}\ }\textbf {\bibinfo {volume} {122}},\ \bibinfo {pages} {113602}
  (\bibinfo {year} {2019})}\BibitemShut {NoStop}%
\bibitem [{\citenamefont {Dudin}\ and\ \citenamefont
  {Kuzmich}(2012)}]{dudin887}%
  \BibitemOpen
  \bibfield  {author} {\bibinfo {author} {\bibfnamefont {Y.}~\bibnamefont
  {Dudin}}\ and\ \bibinfo {author} {\bibfnamefont {A.}~\bibnamefont
  {Kuzmich}},\ }\href@noop {} {\bibfield  {journal} {\bibinfo  {journal}
  {Science}\ }\textbf {\bibinfo {volume} {336}},\ \bibinfo {pages} {887}
  (\bibinfo {year} {2012})}\BibitemShut {NoStop}%
\bibitem [{\citenamefont {Dudin}\ \emph {et~al.}(2012)\citenamefont {Dudin},
  \citenamefont {Li}, \citenamefont {Bariani},\ and\ \citenamefont
  {Kuzmich}}]{dudin2012observation}%
  \BibitemOpen
  \bibfield  {author} {\bibinfo {author} {\bibfnamefont {Y.}~\bibnamefont
  {Dudin}}, \bibinfo {author} {\bibfnamefont {L.}~\bibnamefont {Li}}, \bibinfo
  {author} {\bibfnamefont {F.}~\bibnamefont {Bariani}}, \ and\ \bibinfo
  {author} {\bibfnamefont {A.}~\bibnamefont {Kuzmich}},\ }\href@noop {}
  {\bibfield  {journal} {\bibinfo  {journal} {Nature Physics}\ }\textbf
  {\bibinfo {volume} {8}},\ \bibinfo {pages} {790} (\bibinfo {year}
  {2012})}\BibitemShut {NoStop}%
\bibitem [{\citenamefont {Li}\ \emph {et~al.}(2019{\natexlab{b}})\citenamefont
  {Li}, \citenamefont {Zhou}, \citenamefont {Yang}, \citenamefont {Sun},
  \citenamefont {Liu}, \citenamefont {Bao},\ and\ \citenamefont
  {Pan}}]{li2019semi}%
  \BibitemOpen
  \bibfield  {author} {\bibinfo {author} {\bibfnamefont {J.}~\bibnamefont
  {Li}}, \bibinfo {author} {\bibfnamefont {M.-T.}\ \bibnamefont {Zhou}},
  \bibinfo {author} {\bibfnamefont {C.-W.}\ \bibnamefont {Yang}}, \bibinfo
  {author} {\bibfnamefont {P.-F.}\ \bibnamefont {Sun}}, \bibinfo {author}
  {\bibfnamefont {J.-L.}\ \bibnamefont {Liu}}, \bibinfo {author} {\bibfnamefont
  {X.-H.}\ \bibnamefont {Bao}}, \ and\ \bibinfo {author} {\bibfnamefont
  {J.-W.}\ \bibnamefont {Pan}},\ }\href@noop {} {\bibfield  {journal} {\bibinfo
   {journal} {arXiv preprint arXiv:1903.08902}\ } (\bibinfo {year}
  {2019}{\natexlab{b}})}\BibitemShut {NoStop}%
\bibitem [{\citenamefont {Duan}\ \emph {et~al.}(2001)\citenamefont {Duan},
  \citenamefont {Lukin}, \citenamefont {Cirac},\ and\ \citenamefont
  {Zoller}}]{duan2001long}%
  \BibitemOpen
  \bibfield  {author} {\bibinfo {author} {\bibfnamefont {L.-M.}\ \bibnamefont
  {Duan}}, \bibinfo {author} {\bibfnamefont {M.}~\bibnamefont {Lukin}},
  \bibinfo {author} {\bibfnamefont {J.~I.}\ \bibnamefont {Cirac}}, \ and\
  \bibinfo {author} {\bibfnamefont {P.}~\bibnamefont {Zoller}},\ }\href@noop {}
  {\bibfield  {journal} {\bibinfo  {journal} {Nature}\ }\textbf {\bibinfo
  {volume} {414}},\ \bibinfo {pages} {413} (\bibinfo {year}
  {2001})}\BibitemShut {NoStop}%
\bibitem [{\citenamefont {Zhao}\ \emph {et~al.}(2007)\citenamefont {Zhao},
  \citenamefont {Chen}, \citenamefont {Chen}, \citenamefont {Schmiedmayer},\
  and\ \citenamefont {Pan}}]{zhao2007robust}%
  \BibitemOpen
  \bibfield  {author} {\bibinfo {author} {\bibfnamefont {B.}~\bibnamefont
  {Zhao}}, \bibinfo {author} {\bibfnamefont {Z.-B.}\ \bibnamefont {Chen}},
  \bibinfo {author} {\bibfnamefont {Y.-A.}\ \bibnamefont {Chen}}, \bibinfo
  {author} {\bibfnamefont {J.}~\bibnamefont {Schmiedmayer}}, \ and\ \bibinfo
  {author} {\bibfnamefont {J.-W.}\ \bibnamefont {Pan}},\ }\href@noop {}
  {\bibfield  {journal} {\bibinfo  {journal} {Physical review letters}\
  }\textbf {\bibinfo {volume} {98}},\ \bibinfo {pages} {240502} (\bibinfo
  {year} {2007})}\BibitemShut {NoStop}%
\bibitem [{\citenamefont {Kumar}(1990)}]{kumar1990quantum}%
  \BibitemOpen
  \bibfield  {author} {\bibinfo {author} {\bibfnamefont {P.}~\bibnamefont
  {Kumar}},\ }\href@noop {} {\bibfield  {journal} {\bibinfo  {journal} {Optics
  letters}\ }\textbf {\bibinfo {volume} {15}},\ \bibinfo {pages} {1476}
  (\bibinfo {year} {1990})}\BibitemShut {NoStop}%
\bibitem [{\citenamefont {Huang}\ and\ \citenamefont
  {Kumar}(1992)}]{huang1992observation}%
  \BibitemOpen
  \bibfield  {author} {\bibinfo {author} {\bibfnamefont {J.}~\bibnamefont
  {Huang}}\ and\ \bibinfo {author} {\bibfnamefont {P.}~\bibnamefont {Kumar}},\
  }\href@noop {} {\bibfield  {journal} {\bibinfo  {journal} {Physical review
  letters}\ }\textbf {\bibinfo {volume} {68}},\ \bibinfo {pages} {2153}
  (\bibinfo {year} {1992})}\BibitemShut {NoStop}%
\bibitem [{\citenamefont {Yin}\ \emph {et~al.}(2016)\citenamefont {Yin},
  \citenamefont {Chen}, \citenamefont {Yu}, \citenamefont {Liu}, \citenamefont
  {You}, \citenamefont {Zhou}, \citenamefont {Chen}, \citenamefont {Mao},
  \citenamefont {Huang}, \citenamefont {Zhang} \emph
  {et~al.}}]{yin2016measurement}%
  \BibitemOpen
  \bibfield  {author} {\bibinfo {author} {\bibfnamefont {H.-L.}\ \bibnamefont
  {Yin}}, \bibinfo {author} {\bibfnamefont {T.-Y.}\ \bibnamefont {Chen}},
  \bibinfo {author} {\bibfnamefont {Z.-W.}\ \bibnamefont {Yu}}, \bibinfo
  {author} {\bibfnamefont {H.}~\bibnamefont {Liu}}, \bibinfo {author}
  {\bibfnamefont {L.-X.}\ \bibnamefont {You}}, \bibinfo {author} {\bibfnamefont
  {Y.-H.}\ \bibnamefont {Zhou}}, \bibinfo {author} {\bibfnamefont {S.-J.}\
  \bibnamefont {Chen}}, \bibinfo {author} {\bibfnamefont {Y.}~\bibnamefont
  {Mao}}, \bibinfo {author} {\bibfnamefont {M.-Q.}\ \bibnamefont {Huang}},
  \bibinfo {author} {\bibfnamefont {W.-J.}\ \bibnamefont {Zhang}},  \emph
  {et~al.},\ }\href@noop {} {\bibfield  {journal} {\bibinfo  {journal}
  {Physical review letters}\ }\textbf {\bibinfo {volume} {117}},\ \bibinfo
  {pages} {190501} (\bibinfo {year} {2016})}\BibitemShut {NoStop}%
\bibitem [{\citenamefont {Schuck}\ \emph {et~al.}(2013)\citenamefont {Schuck},
  \citenamefont {Pernice},\ and\ \citenamefont {Tang}}]{schuck2013waveguide}%
  \BibitemOpen
  \bibfield  {author} {\bibinfo {author} {\bibfnamefont {C.}~\bibnamefont
  {Schuck}}, \bibinfo {author} {\bibfnamefont {W.~H.}\ \bibnamefont {Pernice}},
  \ and\ \bibinfo {author} {\bibfnamefont {H.~X.}\ \bibnamefont {Tang}},\
  }\href@noop {} {\bibfield  {journal} {\bibinfo  {journal} {Scientific
  reports}\ }\textbf {\bibinfo {volume} {3}},\ \bibinfo {pages} {1893}
  (\bibinfo {year} {2013})}\BibitemShut {NoStop}%
\bibitem [{\citenamefont {Caspani}\ \emph {et~al.}(2017)\citenamefont
  {Caspani}, \citenamefont {Xiong}, \citenamefont {Eggleton}, \citenamefont
  {Bajoni}, \citenamefont {Liscidini}, \citenamefont {Galli}, \citenamefont
  {Morandotti},\ and\ \citenamefont {Moss}}]{caspani2017integrated}%
  \BibitemOpen
  \bibfield  {author} {\bibinfo {author} {\bibfnamefont {L.}~\bibnamefont
  {Caspani}}, \bibinfo {author} {\bibfnamefont {C.}~\bibnamefont {Xiong}},
  \bibinfo {author} {\bibfnamefont {B.~J.}\ \bibnamefont {Eggleton}}, \bibinfo
  {author} {\bibfnamefont {D.}~\bibnamefont {Bajoni}}, \bibinfo {author}
  {\bibfnamefont {M.}~\bibnamefont {Liscidini}}, \bibinfo {author}
  {\bibfnamefont {M.}~\bibnamefont {Galli}}, \bibinfo {author} {\bibfnamefont
  {R.}~\bibnamefont {Morandotti}}, \ and\ \bibinfo {author} {\bibfnamefont
  {D.~J.}\ \bibnamefont {Moss}},\ }\href@noop {} {\bibfield  {journal}
  {\bibinfo  {journal} {Light: Science \& Applications}\ }\textbf {\bibinfo
  {volume} {6}},\ \bibinfo {pages} {e17100} (\bibinfo {year}
  {2017})}\BibitemShut {NoStop}%
\bibitem [{\citenamefont {Zhong}\ \emph {et~al.}(2018)\citenamefont {Zhong},
  \citenamefont {Li}, \citenamefont {Li}, \citenamefont {Peng}, \citenamefont
  {Su}, \citenamefont {Hu}, \citenamefont {He}, \citenamefont {Ding},
  \citenamefont {Zhang}, \citenamefont {Li}, \citenamefont {Zhang},
  \citenamefont {Wang}, \citenamefont {You}, \citenamefont {Wang},
  \citenamefont {Jiang}, \citenamefont {Li}, \citenamefont {Chen},
  \citenamefont {Liu}, \citenamefont {Lu},\ and\ \citenamefont
  {Pan}}]{PhysRevLett.121.250505}%
  \BibitemOpen
  \bibfield  {author} {\bibinfo {author} {\bibfnamefont {H.-S.}\ \bibnamefont
  {Zhong}}, \bibinfo {author} {\bibfnamefont {Y.}~\bibnamefont {Li}}, \bibinfo
  {author} {\bibfnamefont {W.}~\bibnamefont {Li}}, \bibinfo {author}
  {\bibfnamefont {L.-C.}\ \bibnamefont {Peng}}, \bibinfo {author}
  {\bibfnamefont {Z.-E.}\ \bibnamefont {Su}}, \bibinfo {author} {\bibfnamefont
  {Y.}~\bibnamefont {Hu}}, \bibinfo {author} {\bibfnamefont {Y.-M.}\
  \bibnamefont {He}}, \bibinfo {author} {\bibfnamefont {X.}~\bibnamefont
  {Ding}}, \bibinfo {author} {\bibfnamefont {W.}~\bibnamefont {Zhang}},
  \bibinfo {author} {\bibfnamefont {H.}~\bibnamefont {Li}}, \bibinfo {author}
  {\bibfnamefont {L.}~\bibnamefont {Zhang}}, \bibinfo {author} {\bibfnamefont
  {Z.}~\bibnamefont {Wang}}, \bibinfo {author} {\bibfnamefont {L.}~\bibnamefont
  {You}}, \bibinfo {author} {\bibfnamefont {X.-L.}\ \bibnamefont {Wang}},
  \bibinfo {author} {\bibfnamefont {X.}~\bibnamefont {Jiang}}, \bibinfo
  {author} {\bibfnamefont {L.}~\bibnamefont {Li}}, \bibinfo {author}
  {\bibfnamefont {Y.-A.}\ \bibnamefont {Chen}}, \bibinfo {author}
  {\bibfnamefont {N.-L.}\ \bibnamefont {Liu}}, \bibinfo {author} {\bibfnamefont
  {C.-Y.}\ \bibnamefont {Lu}}, \ and\ \bibinfo {author} {\bibfnamefont {J.-W.}\
  \bibnamefont {Pan}},\ }\href {\doibase 10.1103/PhysRevLett.121.250505}
  {\bibfield  {journal} {\bibinfo  {journal} {Phys. Rev. Lett.}\ }\textbf
  {\bibinfo {volume} {121}},\ \bibinfo {pages} {250505} (\bibinfo {year}
  {2018})}\BibitemShut {NoStop}%
\bibitem [{\citenamefont {Zhong}\ and\ \citenamefont
  {Goldner}(2019)}]{zhong2019emerging}%
  \BibitemOpen
  \bibfield  {author} {\bibinfo {author} {\bibfnamefont {T.}~\bibnamefont
  {Zhong}}\ and\ \bibinfo {author} {\bibfnamefont {P.}~\bibnamefont
  {Goldner}},\ }\href@noop {} {\bibfield  {journal} {\bibinfo  {journal}
  {Nanophotonics}\ } (\bibinfo {year} {2019})}\BibitemShut {NoStop}%
\bibitem [{\citenamefont {Hedges}\ \emph {et~al.}(2010)\citenamefont {Hedges},
  \citenamefont {Longdell}, \citenamefont {Li},\ and\ \citenamefont
  {Sellars}}]{hedges2010efficient}%
  \BibitemOpen
  \bibfield  {author} {\bibinfo {author} {\bibfnamefont {M.~P.}\ \bibnamefont
  {Hedges}}, \bibinfo {author} {\bibfnamefont {J.~J.}\ \bibnamefont
  {Longdell}}, \bibinfo {author} {\bibfnamefont {Y.}~\bibnamefont {Li}}, \ and\
  \bibinfo {author} {\bibfnamefont {M.~J.}\ \bibnamefont {Sellars}},\
  }\href@noop {} {\bibfield  {journal} {\bibinfo  {journal} {Nature}\ }\textbf
  {\bibinfo {volume} {465}},\ \bibinfo {pages} {1052} (\bibinfo {year}
  {2010})}\BibitemShut {NoStop}%
\bibitem [{\citenamefont {Bonarota}\ \emph {et~al.}(2010)\citenamefont
  {Bonarota}, \citenamefont {Ruggiero}, \citenamefont {Le~Gou{\"e}t},\ and\
  \citenamefont {Chaneli{\`e}re}}]{bonarota2010efficiency}%
  \BibitemOpen
  \bibfield  {author} {\bibinfo {author} {\bibfnamefont {M.}~\bibnamefont
  {Bonarota}}, \bibinfo {author} {\bibfnamefont {J.}~\bibnamefont {Ruggiero}},
  \bibinfo {author} {\bibfnamefont {J.-L.}\ \bibnamefont {Le~Gou{\"e}t}}, \
  and\ \bibinfo {author} {\bibfnamefont {T.}~\bibnamefont {Chaneli{\`e}re}},\
  }\href@noop {} {\bibfield  {journal} {\bibinfo  {journal} {Physical Review
  A}\ }\textbf {\bibinfo {volume} {81}},\ \bibinfo {pages} {033803} (\bibinfo
  {year} {2010})}\BibitemShut {NoStop}%
\bibitem [{\citenamefont {Zhou}\ \emph {et~al.}(2012)\citenamefont {Zhou},
  \citenamefont {Lin}, \citenamefont {Yang}, \citenamefont {Li},\ and\
  \citenamefont {Guo}}]{zhou2012realization}%
  \BibitemOpen
  \bibfield  {author} {\bibinfo {author} {\bibfnamefont {Z.-Q.}\ \bibnamefont
  {Zhou}}, \bibinfo {author} {\bibfnamefont {W.-B.}\ \bibnamefont {Lin}},
  \bibinfo {author} {\bibfnamefont {M.}~\bibnamefont {Yang}}, \bibinfo {author}
  {\bibfnamefont {C.-F.}\ \bibnamefont {Li}}, \ and\ \bibinfo {author}
  {\bibfnamefont {G.-C.}\ \bibnamefont {Guo}},\ }\href@noop {} {\bibfield
  {journal} {\bibinfo  {journal} {Physical review letters}\ }\textbf {\bibinfo
  {volume} {108}},\ \bibinfo {pages} {190505} (\bibinfo {year}
  {2012})}\BibitemShut {NoStop}%
\bibitem [{\citenamefont {G{\"u}ndo{\u{g}}an}\ \emph
  {et~al.}(2012)\citenamefont {G{\"u}ndo{\u{g}}an}, \citenamefont {Ledingham},
  \citenamefont {Almasi}, \citenamefont {Cristiani},\ and\ \citenamefont
  {de~Riedmatten}}]{gundougan2012quantum}%
  \BibitemOpen
  \bibfield  {author} {\bibinfo {author} {\bibfnamefont {M.}~\bibnamefont
  {G{\"u}ndo{\u{g}}an}}, \bibinfo {author} {\bibfnamefont {P.~M.}\ \bibnamefont
  {Ledingham}}, \bibinfo {author} {\bibfnamefont {A.}~\bibnamefont {Almasi}},
  \bibinfo {author} {\bibfnamefont {M.}~\bibnamefont {Cristiani}}, \ and\
  \bibinfo {author} {\bibfnamefont {H.}~\bibnamefont {de~Riedmatten}},\
  }\href@noop {} {\bibfield  {journal} {\bibinfo  {journal} {Physical review
  letters}\ }\textbf {\bibinfo {volume} {108}},\ \bibinfo {pages} {190504}
  (\bibinfo {year} {2012})}\BibitemShut {NoStop}%
\bibitem [{\citenamefont {Afzelius}\ \emph {et~al.}(2009)\citenamefont
  {Afzelius}, \citenamefont {Simon}, \citenamefont {de~Riedmatten},\ and\
  \citenamefont {Gisin}}]{PhysRevA.79.052329}%
  \BibitemOpen
  \bibfield  {author} {\bibinfo {author} {\bibfnamefont {M.}~\bibnamefont
  {Afzelius}}, \bibinfo {author} {\bibfnamefont {C.}~\bibnamefont {Simon}},
  \bibinfo {author} {\bibfnamefont {H.}~\bibnamefont {de~Riedmatten}}, \ and\
  \bibinfo {author} {\bibfnamefont {N.}~\bibnamefont {Gisin}},\ }\href
  {\doibase 10.1103/PhysRevA.79.052329} {\bibfield  {journal} {\bibinfo
  {journal} {Phys. Rev. A}\ }\textbf {\bibinfo {volume} {79}},\ \bibinfo
  {pages} {052329} (\bibinfo {year} {2009})}\BibitemShut {NoStop}%
\bibitem [{\citenamefont {Afzelius}\ and\ \citenamefont
  {Simon}(2010)}]{afzelius2010impedance}%
  \BibitemOpen
  \bibfield  {author} {\bibinfo {author} {\bibfnamefont {M.}~\bibnamefont
  {Afzelius}}\ and\ \bibinfo {author} {\bibfnamefont {C.}~\bibnamefont
  {Simon}},\ }\href@noop {} {\bibfield  {journal} {\bibinfo  {journal}
  {Physical Review A}\ }\textbf {\bibinfo {volume} {82}},\ \bibinfo {pages}
  {022310} (\bibinfo {year} {2010})}\BibitemShut {NoStop}%
\bibitem [{\citenamefont {Saffman}\ \emph {et~al.}(2010)\citenamefont
  {Saffman}, \citenamefont {Walker},\ and\ \citenamefont
  {M{\o}lmer}}]{saffman2010quantum}%
  \BibitemOpen
  \bibfield  {author} {\bibinfo {author} {\bibfnamefont {M.}~\bibnamefont
  {Saffman}}, \bibinfo {author} {\bibfnamefont {T.~G.}\ \bibnamefont {Walker}},
  \ and\ \bibinfo {author} {\bibfnamefont {K.}~\bibnamefont {M{\o}lmer}},\
  }\href@noop {} {\bibfield  {journal} {\bibinfo  {journal} {Reviews of Modern
  Physics}\ }\textbf {\bibinfo {volume} {82}},\ \bibinfo {pages} {2313}
  (\bibinfo {year} {2010})}\BibitemShut {NoStop}%
\bibitem [{\citenamefont {Lodahl}\ \emph {et~al.}(2015)\citenamefont {Lodahl},
  \citenamefont {Mahmoodian},\ and\ \citenamefont
  {Stobbe}}]{lodahl2015interfacing}%
  \BibitemOpen
  \bibfield  {author} {\bibinfo {author} {\bibfnamefont {P.}~\bibnamefont
  {Lodahl}}, \bibinfo {author} {\bibfnamefont {S.}~\bibnamefont {Mahmoodian}},
  \ and\ \bibinfo {author} {\bibfnamefont {S.}~\bibnamefont {Stobbe}},\
  }\href@noop {} {\bibfield  {journal} {\bibinfo  {journal} {Reviews of Modern
  Physics}\ }\textbf {\bibinfo {volume} {87}},\ \bibinfo {pages} {347}
  (\bibinfo {year} {2015})}\BibitemShut {NoStop}%
\bibitem [{\citenamefont {Friedler}\ \emph {et~al.}(2009)\citenamefont
  {Friedler}, \citenamefont {Sauvan}, \citenamefont {Hugonin}, \citenamefont
  {Lalanne}, \citenamefont {Claudon},\ and\ \citenamefont
  {G{\'e}rard}}]{friedler2009solid}%
  \BibitemOpen
  \bibfield  {author} {\bibinfo {author} {\bibfnamefont {I.}~\bibnamefont
  {Friedler}}, \bibinfo {author} {\bibfnamefont {C.}~\bibnamefont {Sauvan}},
  \bibinfo {author} {\bibfnamefont {J.-P.}\ \bibnamefont {Hugonin}}, \bibinfo
  {author} {\bibfnamefont {P.}~\bibnamefont {Lalanne}}, \bibinfo {author}
  {\bibfnamefont {J.}~\bibnamefont {Claudon}}, \ and\ \bibinfo {author}
  {\bibfnamefont {J.-M.}\ \bibnamefont {G{\'e}rard}},\ }\href@noop {}
  {\bibfield  {journal} {\bibinfo  {journal} {Optics express}\ }\textbf
  {\bibinfo {volume} {17}},\ \bibinfo {pages} {2095} (\bibinfo {year}
  {2009})}\BibitemShut {NoStop}%
\bibitem [{\citenamefont {Claudon}\ \emph {et~al.}(2010)\citenamefont
  {Claudon}, \citenamefont {Bleuse}, \citenamefont {Malik}, \citenamefont
  {Bazin}, \citenamefont {Jaffrennou}, \citenamefont {Gregersen}, \citenamefont
  {Sauvan}, \citenamefont {Lalanne},\ and\ \citenamefont
  {G{\'e}rard}}]{claudon2010highly}%
  \BibitemOpen
  \bibfield  {author} {\bibinfo {author} {\bibfnamefont {J.}~\bibnamefont
  {Claudon}}, \bibinfo {author} {\bibfnamefont {J.}~\bibnamefont {Bleuse}},
  \bibinfo {author} {\bibfnamefont {N.~S.}\ \bibnamefont {Malik}}, \bibinfo
  {author} {\bibfnamefont {M.}~\bibnamefont {Bazin}}, \bibinfo {author}
  {\bibfnamefont {P.}~\bibnamefont {Jaffrennou}}, \bibinfo {author}
  {\bibfnamefont {N.}~\bibnamefont {Gregersen}}, \bibinfo {author}
  {\bibfnamefont {C.}~\bibnamefont {Sauvan}}, \bibinfo {author} {\bibfnamefont
  {P.}~\bibnamefont {Lalanne}}, \ and\ \bibinfo {author} {\bibfnamefont
  {J.-M.}\ \bibnamefont {G{\'e}rard}},\ }\href@noop {} {\bibfield  {journal}
  {\bibinfo  {journal} {Nature Photonics}\ }\textbf {\bibinfo {volume} {4}},\
  \bibinfo {pages} {174} (\bibinfo {year} {2010})}\BibitemShut {NoStop}%
\bibitem [{\citenamefont {Versteegh}\ \emph {et~al.}(2014)\citenamefont
  {Versteegh}, \citenamefont {Reimer}, \citenamefont {J{\"o}ns}, \citenamefont
  {Dalacu}, \citenamefont {Poole}, \citenamefont {Gulinatti}, \citenamefont
  {Giudice},\ and\ \citenamefont {Zwiller}}]{versteegh2014observation}%
  \BibitemOpen
  \bibfield  {author} {\bibinfo {author} {\bibfnamefont {M.~A.}\ \bibnamefont
  {Versteegh}}, \bibinfo {author} {\bibfnamefont {M.~E.}\ \bibnamefont
  {Reimer}}, \bibinfo {author} {\bibfnamefont {K.~D.}\ \bibnamefont
  {J{\"o}ns}}, \bibinfo {author} {\bibfnamefont {D.}~\bibnamefont {Dalacu}},
  \bibinfo {author} {\bibfnamefont {P.~J.}\ \bibnamefont {Poole}}, \bibinfo
  {author} {\bibfnamefont {A.}~\bibnamefont {Gulinatti}}, \bibinfo {author}
  {\bibfnamefont {A.}~\bibnamefont {Giudice}}, \ and\ \bibinfo {author}
  {\bibfnamefont {V.}~\bibnamefont {Zwiller}},\ }\href@noop {} {\bibfield
  {journal} {\bibinfo  {journal} {Nature communications}\ }\textbf {\bibinfo
  {volume} {5}},\ \bibinfo {pages} {5298} (\bibinfo {year} {2014})}\BibitemShut
  {NoStop}%
\bibitem [{\citenamefont {Huber}\ \emph {et~al.}(2014)\citenamefont {Huber},
  \citenamefont {Predojevic}, \citenamefont {Khoshnegar}, \citenamefont
  {Dalacu}, \citenamefont {Poole}, \citenamefont {Majedi},\ and\ \citenamefont
  {Weihs}}]{huber2014polarization}%
  \BibitemOpen
  \bibfield  {author} {\bibinfo {author} {\bibfnamefont {T.}~\bibnamefont
  {Huber}}, \bibinfo {author} {\bibfnamefont {A.}~\bibnamefont {Predojevic}},
  \bibinfo {author} {\bibfnamefont {M.}~\bibnamefont {Khoshnegar}}, \bibinfo
  {author} {\bibfnamefont {D.}~\bibnamefont {Dalacu}}, \bibinfo {author}
  {\bibfnamefont {P.~J.}\ \bibnamefont {Poole}}, \bibinfo {author}
  {\bibfnamefont {H.}~\bibnamefont {Majedi}}, \ and\ \bibinfo {author}
  {\bibfnamefont {G.}~\bibnamefont {Weihs}},\ }\href@noop {} {\bibfield
  {journal} {\bibinfo  {journal} {Nano Letters}\ }\textbf {\bibinfo {volume}
  {14}},\ \bibinfo {pages} {7107} (\bibinfo {year} {2014})}\BibitemShut
  {NoStop}%
\bibitem [{\citenamefont {J{\"o}ns}\ \emph {et~al.}(2017)\citenamefont
  {J{\"o}ns}, \citenamefont {Schweickert}, \citenamefont {Versteegh},
  \citenamefont {Dalacu}, \citenamefont {Poole}, \citenamefont {Gulinatti},
  \citenamefont {Giudice}, \citenamefont {Zwiller},\ and\ \citenamefont
  {Reimer}}]{jons2017bright}%
  \BibitemOpen
  \bibfield  {author} {\bibinfo {author} {\bibfnamefont {K.~D.}\ \bibnamefont
  {J{\"o}ns}}, \bibinfo {author} {\bibfnamefont {L.}~\bibnamefont
  {Schweickert}}, \bibinfo {author} {\bibfnamefont {M.~A.}\ \bibnamefont
  {Versteegh}}, \bibinfo {author} {\bibfnamefont {D.}~\bibnamefont {Dalacu}},
  \bibinfo {author} {\bibfnamefont {P.~J.}\ \bibnamefont {Poole}}, \bibinfo
  {author} {\bibfnamefont {A.}~\bibnamefont {Gulinatti}}, \bibinfo {author}
  {\bibfnamefont {A.}~\bibnamefont {Giudice}}, \bibinfo {author} {\bibfnamefont
  {V.}~\bibnamefont {Zwiller}}, \ and\ \bibinfo {author} {\bibfnamefont
  {M.~E.}\ \bibnamefont {Reimer}},\ }\href@noop {} {\bibfield  {journal}
  {\bibinfo  {journal} {Scientific reports}\ }\textbf {\bibinfo {volume} {7}},\
  \bibinfo {pages} {1700} (\bibinfo {year} {2017})}\BibitemShut {NoStop}%
\end{thebibliography}%
    
\end{document}